\documentclass[12pt]{article}
\usepackage{amssymb}
\usepackage{epsfig}
\usepackage{amssymb,amsmath}
\usepackage{slashed}

\newcommand{\bea}{\begin{eqnarray}}
\newcommand{\eea}{\end{eqnarray}}

 \makeatletter
\def\alt{\mathrel{\mathpalette\gl@align<}}
\def\agt{\mathrel{\mathpalette\gl@align>}}
\def\gl@align#1#2{\lower.6ex\vbox{\baselineskip\z@skip\lineskip\z@
\ialign{$\m@th#1\hfil##\hfil$\crcr#2\crcr\sim\crcr}}} \makeatother

\begin{document}
\begin{flushright}
BA-08-15\\
OSU-HEP-08-06\\
\end{flushright}
\vspace*{1.0cm}

\begin{center}
\baselineskip 20pt {\Large\bf Higgs Boson Mass, Sparticle Spectrum
and Little Hierarchy Problem in Extended MSSM} \vspace{1cm}

{\large K.S. Babu$^{a}$, Ilia Gogoladze$^{b,}$\footnote{ On  leave
of absence from: Andronikashvili Institute of Physics, GAS, 380077
Tbilisi, Georgia. \\ \hspace*{0.5cm} }, Mansoor Ur Rehman$^{b}$ and
Qaisar Shafi$^{b}$} \vspace{.5cm}

{\baselineskip 20pt \it $^a$ Department of Physics, Oklahoma State
University, Stillwater,
OK 74078, USA \\
$^b$Bartol Research Institute, Department of Physics and Astronomy, \\
University of Delaware, Newark, DE 19716, USA \\

 }
\vspace{.5cm}

\end{center}

\begin{abstract}

We investigate the impact of TeV-scale matter belonging to 
complete vectorlike multiplets of unified groups on the 
lightest Higgs boson in the MSSM. We find that consistent
with perturbative unification  and electroweak precision data the
mass $m_h$ can be as large as $160$ GeV. These extended MSSM models 
can also render the little hierarchy problem less severe, but only
for lower values of $m_h (\lesssim 125)$ GeV. We present estimates
for the sparticle mass spectrum in these models.

\end{abstract}

\date{}

\section{Introduction}

The LEP2 lower bound $m_{h}\geq $ $114.4$ GeV \cite{LEP2} on the
Standard Model (SM) Higgs boson mass poses a significant challenge
for the minimal
supersymmetric standard model (MSSM). With the tree level upper bound of $%
M_{Z}$ on the mass of the (lightest) SM-like Higgs boson in MSSM,
significant radiative corrections are required to lift this mass
above the LEP2 bound. This situation has been further exacerbated by
the most recently quoted value of $172.6\pm 1.4$ GeV for the top
quark pole mass \cite{Group:2008nq}, significantly lower from
earlier values which not so long ago were closer to $176$ GeV and
higher \cite{b2}. With radiative corrections proportional to the
fourth power of $m_{t}$, this leads to a reduced value for $m_{h}$
unless the magnitude of some MSSM parameters such as the stop mass $m_{%
\widetilde{t}}$ (or $M_{S}$) and the soft trilinear parameter
$A_{t}$ are suitably increased. Values of $m_{h}$ of around $123$
GeV or so require stop masses as well as $\left| A_{t}\right| $
close to the TeV scale or higher. Such large values, in turn, lead
to the so-called {\it {little hierarchy}} problem \cite{b5} because,
when dealing with radiative electroweak symmetry breaking, TeV scale
quantities must conspire to yield the electroweak mass scale
$M_{Z}$.

\begin{figure}[t]
\centering \includegraphics[angle=0, width=13cm]{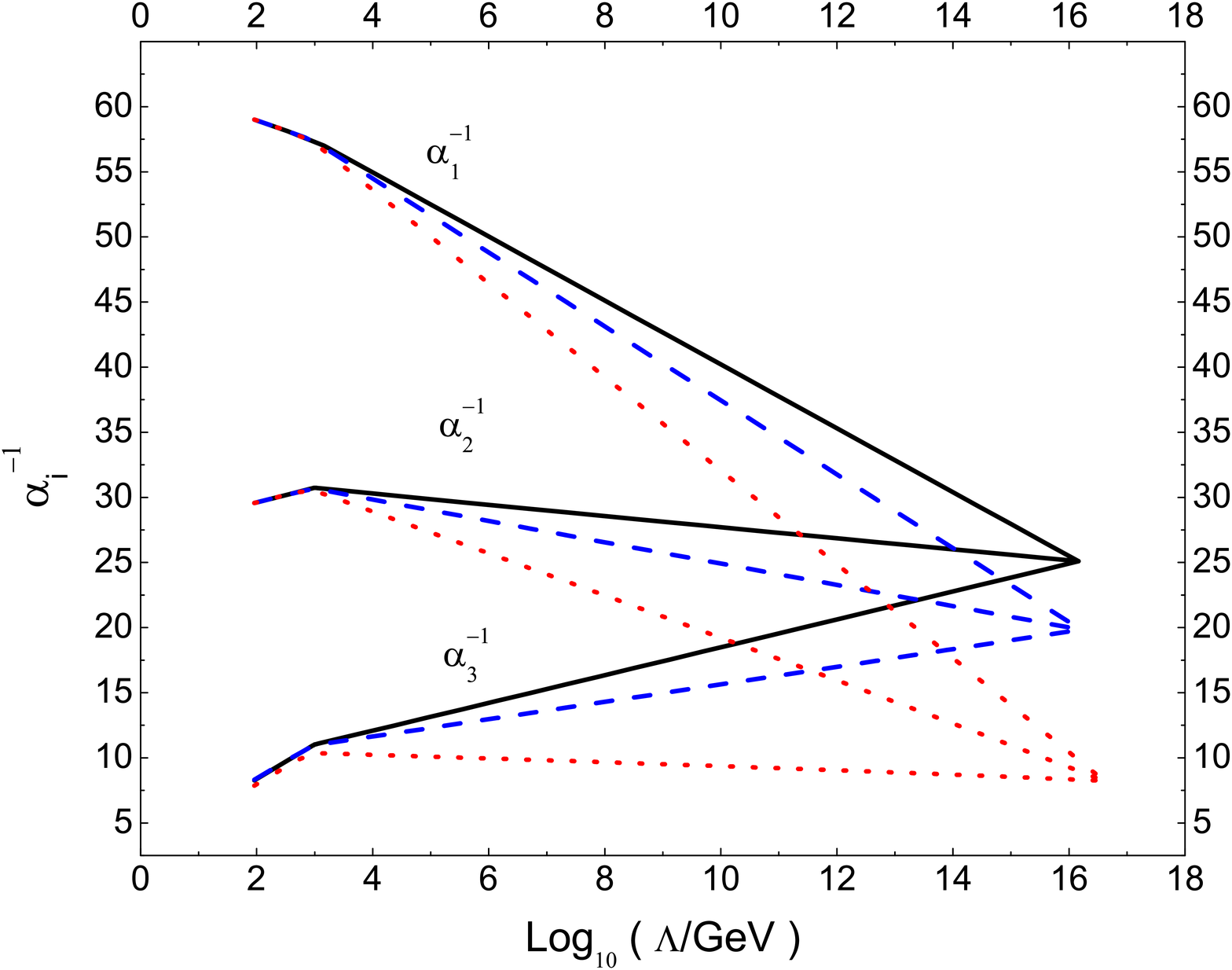}
\vspace{-1cm} \caption{ Gauge coupling evolution with the effective
SUSY breaking scale $M_{S}= 1$ TeV and tan$\protect\beta $ =
10. Solid lines correspond to MSSM. Dashed lines
correspond to MSSM$+5+\bar{5}$. Dotted lines are for
MSSM $+10+\overline{10}$. The vectorlike
masses for all these cases are set equal to $500$ GeV.} \label{gu}
\end{figure}

In this paper we address these two related conundrums of the MSSM by
introducing TeV scale vectorlike matter superfields which reside in
complete SU(5) or SO(10) multiplets. Such complete multiplets, it is well
known, do not spoil unification of the MSSM gauge couplings.
We illustrate this in Figure \ref{gu}, where the gauge coupling evolution is
compared, using two loop RGEs, for the case of MSSM plus
complete multiplets $10 + \overline{10}$ and $5 + \overline{5}$ of $SU(5)$.
If these vectorlike matter fields do not acquire Planck-scale
masses, it appears quite plausible that they will end up order
TeV masses. $R$-symmetries, for instance, can
forbid Planck scale masses, but allow TeV scale masses proportional to
the SUSY breaking scale. The Higgs(ino) mass term (the $\mu$ parameter)
for the $H_u-H_d$ superfields
is an example where this happens already in the MSSM \cite{Giudice:1988yz}.
For the vectorlike matter to have any significant effect on the
`upper' bounds on $m_h$, it is crucial that they have masses of order TeV,
otherwise their effects on $m_h$ will decouple.

 In studying this possibility of extending the MSSM, we employ the
perturbativity and GUT unification constraints. It turns out that
perturbative unification can be maintained if we introduce either
  ($i$) one pair of ($10+\overline{10}$), or ($ii$%
) up to four pairs of ($5+\overline{5}$), or ($iii$) one set of ($10+%
\overline{10}+5+\overline{5}$) \cite{Babu:1996zv}, where the representations
refer to $SU(5)$. In addition, any number of SM singlet fileds are also
allowed. Some particles in
these new supermultiplets couple to the MSSM Higgs doublet $H_{u}$,
and with masses of order $0.5-1$ TeV, their radiative contributions
alone can lift $m_{h}$ to values as high as $160$ GeV. This is
achieved without requiring the standard  MSSM sparticles to be much
heavier than their present experimental lower bounds. We explore the
impact of the additional multiplets on the MSSM parameter space, and
obtain the low energy sparticle spectrum in the mSUGRA framework. A
comparison is presented, using semi-analytic estimates, between the
minimal and extended MSSM sparticle spectrum. The impact of these
new particles on the little hierarchy problem is also discussed.

With the inclusion of these new vectorlike particles, the lightest
Higgs boson mass, as previously noted, can be significantly
increased. However, we find that resolving the little hierarchy
problem is somewhat more tricky. The new Yukawa couplings of $H_u$
to vectorlike matter, which helps in raising $m_h$, also has the
effect of raising the soft Higgs mass parameter $m_{H_{u}}^2$, which
could exacerbate the little hierarchy problem. If the new Yukawa
couplings of $H_u$ are relatively small, the little hierarchy
problem improves relative to the MSSM, since the cumulative effect
of the top Yukawa coupling $y_t$ on $m_{H_{u}}^2$ becomes smaller
than in MSSM. This comes about since $y_t$ has a smaller value at
the GUT scale ($y_t \sim 0.15$) compared to the MSSM case $(y_t \sim
0.5)$. Thus we identify two regions of the parameter space as being
of special interest: one where the little hierarchy problem becomes
worse than in MSSM, but where $m_h \sim 130-160$ GeV can be achieved
, and another where the little hierarchy problem is relaxed, but
where $m_h \lesssim 125$ GeV. The latter possibility appears to
us to be quite interesting, as it assumes MSSM sparticle masses to
be moderate, of order $200  - 500$ GeV.

\section{New vectorlike matter and precision constraints}

It is well known that one can extend the matter sector of the MSSM
and still preserve the beautiful result of gauge coupling
unification provided that the additional matter superfields fall
into complete multiplets of any unified group which contains the SM,
such as SU(5). Such extended scenarios
with TeV scale matter multiplets are well motivated. Within string 
theory, for instance, one often
finds light (TeV scale) multiplets in the spectrum
\cite{Dienes:1996du}, and even within the framework of GUTs one can
find extra complete multiplets with masses around the TeV
scale~\cite{Babu:1996zv}.

An important constraint on GUT representations and how many there
can be at low ($\sim$ TeV) scale comes from the perturbativity condition,
which requires that the three MSSM gauge couplings remain
perturbative up to $M_{G}$.
One finds that there are several choices
to satisfy this constraint: {\it (i)} one pair of
$(10+\overline{10})$, {\it (ii)} up to 4 pairs of $(5+ \bar5)$'s, or
{\it (iii)} the combination, $(5+\bar 5+10+\overline{10})$. Here 
all representations refer to multiplets of $SU(5)$.
In
addition, any number of MSSM gauge singlets can be added without
sacrificing unification or perturbativity. Option (iii), along
with a pair of MSSM gauge singlets fits nicely in $SO(10)$ models.

Cases (i) and (iii) have been studied before in the literature. For
example, the authors in \cite{Moroi:1992zk} conclude that the mass
of the lightest CP even Higgs mass could be pushed up to 180~GeV,
consistent with all perturbativity constraints.
When updated to account for the recent electroweak precision data, specifically the
$T$ parameter, and the current value of the top quark mass, and improved to include two--loop
RGE effects and finite corrections to the Higgs boson mass, we find that these
scenarios admit $m_h$ only as large as $160$ GeV, which is significantly
smaller than the bounds in \cite{Moroi:1992zk}.

It is clear that new matter will contribute at one loop level  to
CP-even Higgs mass if there is direct coupling among new matter and
the MSSM Higgs field.  In case (i), a new couplings
$10\cdot 10\cdot H_u$ is allowed, analogous to the top--quark Yukawa
coupling, but involving the charge $2/3$ quark from the $10$-plet.
(Here we use for simplicity $SU(5)$ notation, but with the
understanding that $H_u$ and $H_d$ are not complete multiplets of
$SU(5)$.) This new Yukawa coupling can modify the upper limit on
$m_h$, which we will study in detail, taking into account
perturbativity constraints. By itself, case (ii) does not allow for
any new Yukawa coupling unless the new states in the $\overline{5}$
are mixed with the usual $d^c$-quarks and lepton doublets. Such a 
possibility is even more
strongly constrained (by flavor violation and unitarity of the CKM
matrix, among others), and so we will forbid all such mixing.  Once
we add gauge singlets $1$, couplings such as $\bar 5 H_u 1$ are
allowed (Only the lepton--like doublets from
$\bar{5}$ will be involved in this Yukawa coupling.) We will analyze
the effects of such couplings on $m_h$ in detail.  Case (iii) is a
combination of (i) and (ii), which will also be studied in detail.

There are constraints on the couplings and masses of new matter
fields. Most important are the
constraints from the $S$ and $T$ parameters which limit the number
of additional {\it chiral}\/ generations. Consistent with these
constraints, one should add new matter which is predominantly
vectorlike.

In the limit where the vectorlike mass is much heavier than the
chiral mass term (mass term arising from Yukawa coupling to the 
Higgs doublets), the contribution to the $T$ parameter from a
single chiral fermion is approximately~\cite{Lavoura:1992np}:
\begin{equation}
\delta T=\frac{ N (\kappa v)^2}{10 \pi \sin^2\theta_W m^2_W}\left[ \left(
\frac{\kappa v}{M_V}\right)^2  +O\left( \frac{\kappa
v}{M_V}\right)^4\right], \label{tpar}
\end{equation}
where $\kappa$ is the new chiral Yukawa coupling, $v$ is
VEV of the corresponding Higgs field, and $N$ counts the additional
number of SU(2) doublets. For instance, $N=3$ when $10 + \overline
{10}$ is considered at low scale, while  $N=1$ for the $5+\bar 5$
case. From precision electroweak data $T\leq 0.06(0.14)$ at 95\% CL for
$m_h=117$ GeV ($300$ GeV) ~\cite{PDG}. We
will take $\delta T<0.1$ as a realistic bound and apply it in our analysis.
We then see from Eq.~(\ref{tpar}) that with $M_V$ around 1~TeV, the
Yukawa coupling $\kappa$ can be $O(1)$.

\section{Higgs mass bound}

\subsection{{MSSM }$\mathbf{+}$ $10$ $\mathbf{+}$ $\overline{10}$}

The representation  $10+\overline{10}$ of SU(5)
decomposes under the MSSM gauge symmetry as follows:%
\begin{eqnarray}
10+\overline{10} &=&Q_{10}\left( 3,2,\frac{1}{6}\right) +%
\overline{Q}_{10}\left( \overline{3},2,-\frac{1}{6}\right) +U_{10}\left(
\overline{3},1,-\frac{2}{3}\right) +\overline{U}_{10}\left( 3,1,\frac{2}{3}%
\right)  \nonumber \\
&&+E_{10}\left( 1,1,1\right) +\overline{E}_{10}\left( 1,1,-1\right).
\end{eqnarray}%
We assume for the vectorlike matter  $10+\overline{10}$ the same $R$
parity as the MSSM Higgs chiral superfields. So there is no mixing
of this new matter with quarks, but they couple to the Higgs
doublets. The contribution to the superpotential from these
couplings is
\begin{equation}
W=\kappa _{10}Q_{10}U_{10}H_{u}+\kappa _{10}^{\prime }\overline{Q}_{10}%
\overline{U}_{10}H_{d}+M_{V}\left( \overline{Q}_{10}Q_{10}+\overline{U}%
_{10}U_{10}+\overline{E}_{10}E_{10}\right), \label{nn1}
\end{equation}%
where, for simplicity, we have taken a common vectorlike mass (at the
GUT scale $M_G$). Thus the up quark-like pieces of the $10$ and
$\overline{10}$ get Dirac {\it and}\/ vectorlike masses, while
leaving the $E_{10}$-lepton-like pieces with only vectorlike
masses.
 We assume that
$\kappa _{10}\gg \kappa _{10}^{\prime }$ because the contribution
coming from the coupling $ \kappa _{10}^{\prime }$  reduces the
light higgs mass similar to what we have with bottom Yukawa
contribution which becomes prominent for large tan$\beta $
\cite{Brignole}.

\begin{figure}[t]
\centering \includegraphics[angle=0, width=13cm]{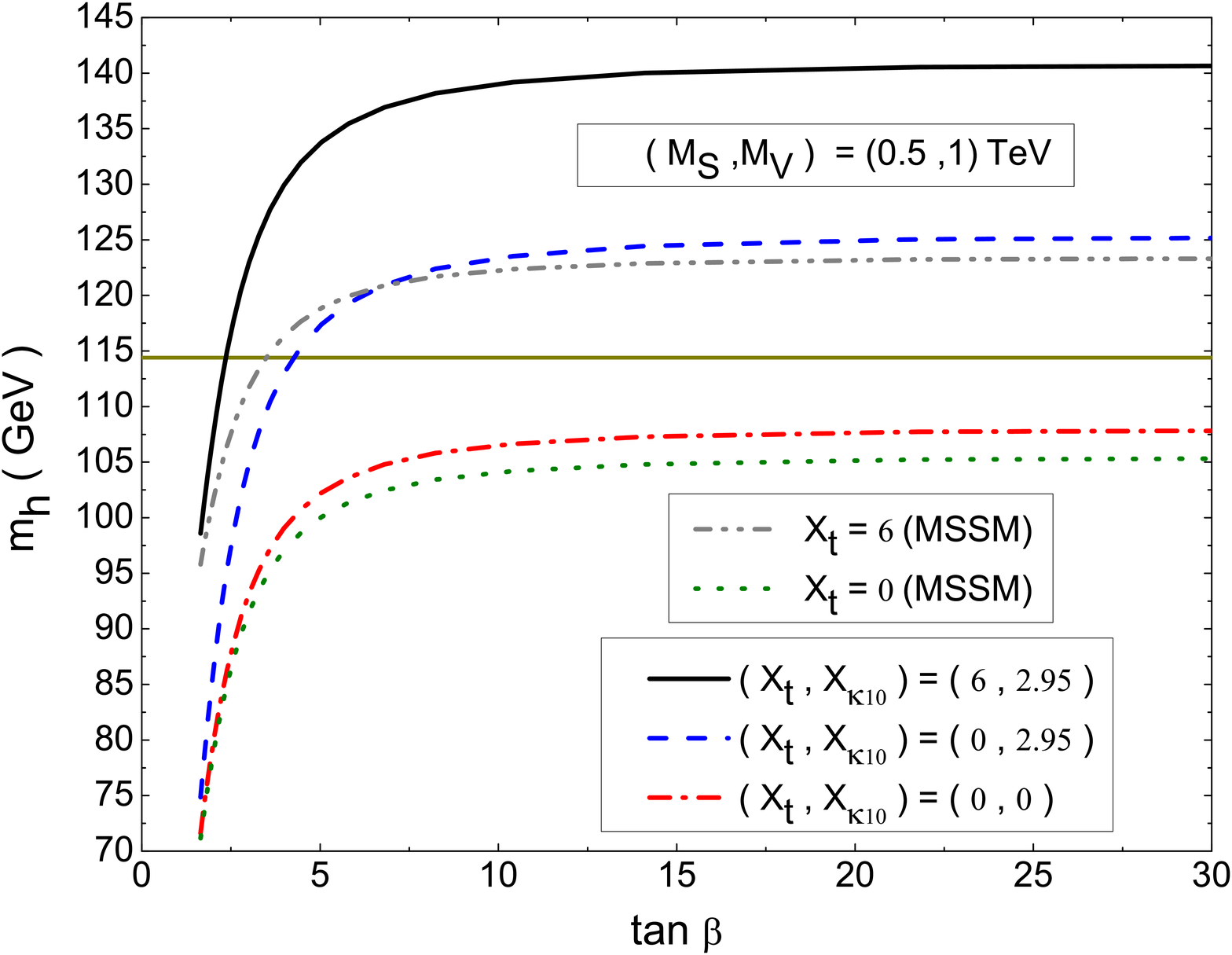}
\vspace{-1cm} \caption{Upper bounds for the lightest CP-even Higgs boson mass vs
$\tan\beta$, for different maximal and minimal values of $X_t$,
$X_{\kappa_{10}}$, with $M_S=500$ GeV, $M_V=1$ TeV and $M_t=172.6$
GeV.  Dotted  line corresponds to MSSM ($X_t=0$).
Dashed-double dotted line describes MSSM with ($X_t=6$).
Dashed-dotted curve is for MSSM $+10+\overline{10}$.
$\kappa_{10} \approx 2$ at $M_{G}$.
  Dashed line shows Higgs mass with
 $X_{10}=2.95$ and $X_t=0$. Solid line corresponds to $X_{10}=2.95$ and $X_t=6$.
 The solid horizontal line denotes the LEP2 bound $m_h=114.4$ GeV.} \label{mh101}
\end{figure}

Employing the effective potential approach we calculate the
additional contribution from the vectorlike particles to the CP even
Higgs mass at one loop level. A similar calculation was carried out
in ref. \cite{ b3}.
\begin{eqnarray}
\left[ m_{h}^{2}\right] _{10} &=&-M_{Z}^{2}\cos ^{2}2\beta \left(
\frac{3}{8\pi ^{2}}\kappa_{10} ^{2}t_{V}\right)
+\frac{3}{4\pi ^{2}}\kappa_{10} ^{4}v^{2}\sin ^{2}\beta \left[ t_{V}+\frac{1%
}{2}X_{\kappa_{10} }\right],  \label{eq1}
\end{eqnarray}%
 where we have assumed $M_{V}\gg M_{D}$. The corrected
 expression for
$X_{\kappa_{10} }$ (compare result in ref. \cite{Moroi:1992zk})
 is given as follows%
\begin{equation}
X_{\kappa_{10} }=\frac{4\widetilde{A}_{\kappa_{10} }^{2}\left(
3M_{S}^{2}+2M_{V}^{2}\right) -\widetilde{A}_{\kappa_{10}
}^{4}-8M_{S}^{2}M_{V}^{2}-10M_{S}^{4}}{6\left(
M_{S}^{2}+M_{V}^{2}\right) ^{2}} \label{f1}
\end{equation}%
and
\begin{equation}
t_{V}=\log \left( \frac{%
M_{S}^{2}+M_{V}^{2}}{M_{V}^{2}}\right), \label{mm7}
\end{equation}
 where $\widetilde{A}_{\kappa_{10}}=A_{\kappa_{10}}-\mu \cot \beta $, 
$A_{\kappa_{10}}$ is the $Q_{10}-U_{10}$ soft mixing
parameter and $\mu$ is the MSSM Higgs bilinear mixing term.
$M_S\simeq \sqrt{m_{\tilde{Q}_{3}}\,m_{\tilde{U}_{3}^c}}$, where
$m_{\tilde{Q}_{3}}$ and $m_{\tilde{U}_{3}^c}$ are the stop left and
stop right soft SUSY breaking masses at low scale.

For completeness we present the leading  1- and 2- loop
contributions to the CP-even Higgs boson mass in the MSSM
\cite{at, Carena:1995wu}%
\begin{eqnarray}
\left[ m_{h}^{2}\right] _{MSSM} &=&M_{Z}^{2}\cos ^{2}2\beta \left( 1-\frac{3%
}{8\pi ^{2}}\frac{m_{t}^{2}}{v^{2}}t\right)  \nonumber \\
&+&\frac{3}{4\pi ^{2}}\frac{m_{t}^{4}}{v^{2}}\left[ t+\frac{1}{2}X_{t}+\frac{%
1}{\left( 4\pi \right) ^{2}}\left(
\frac{3}{2}\frac{m_{t}^{2}}{v^{2}}-32\pi \alpha _{s}\right) \left(
X_{t}t+t^{2}\right) \right],  \label{eq2}
\end{eqnarray}
where
\\
\begin{eqnarray}
t =\log \left(
\frac{M_{S}^{2}}{M_{t}^{2}}\right),~
X_{t} &=&\frac{2\widetilde{A}_{t}^{2}}{M_{S}^{2}}\left( 1-\frac{\widetilde{A}%
_{t}^{2}}{12M_{S}^{2}}\right). \label{A1}
\end{eqnarray}%
Also $\widetilde{A}_{t}=A_{t}-\mu \cot \beta $, where
$A_{t}$ denotes the stop left and stop right soft
mixing parameter.

In our model for the light Higgs mass we have
\begin{eqnarray}
m_h^2= \left[ m_{h}^{2}\right] _{MSSM} + \left[ m_{h}^{2}\right]
_{10}. \label{max_10}
\end{eqnarray}%
From Eq. (\ref{eq1}) we can see that the Higgs mass is very
sensitive to the value of $\kappa_{10}$, which we cannot take to be
arbitrary large because the theory should be perturbative up to GUT
scale. We therefore should solve the following RGE for $\kappa_{10}$
to make sure that it remains perturbative up to the GUT scale:
\begin{eqnarray}
 \frac{d \kappa_{10}}{dt}&=&\frac{\kappa_{10}}{2(4 \pi)^2}  \left( \left(
\frac{16}{3}g^2_{3}+3 g^2_{2}+\frac{13}{15}g^2_{1}
-6\kappa_{10}^{2}-3y_{t}^2\right) \right.     \nonumber \\
&-&\frac{1}{(4 \pi)^2}\left(\frac{3913}{450}g^4_{1}+\frac{33}{2}g^4_{2}+\frac{128}{9}g^4_{3}+g^2_{1}g^2_{2}+\frac{136}{45}g^2_{1}g^2_{3}+8g^2_{2}g^2_{3}+
\left(\frac{2}{5}g^2_{1}+6g^2_{2} \right)\kappa_{10}^2\right.   \nonumber \\
&+& \left. \left.  \left(\frac{4}{5}g^2_{1}+16g^2_{3} \right) \left( y^2_{t}+\kappa ^2_{10}\right) -
9\left( y^4_{t}+\kappa ^4_{10}\right)-9 \kappa_{10}^2\left( y^2_{t}+\kappa_{10}^2\right)-4 \kappa_{10}^4 \right) \right),
\end{eqnarray}%
where $g_3$, $g_2$ and $g_1$ are strong, weak and hypercharge gauge
couplings respectively and $y_t$ denotes the top Yukawa coupling.
Because the new matter couples to $H_u$ (see Eq. (\ref{nn1})) there
are additional contributions to the RGE for $y_t$ at two loop level:
\begin{eqnarray}
 \frac{d y_t}{dt}&=&\frac{y_t}{2(4 \pi)^2}  \left( \left(
\frac{16}{3}g^2_{3}+3 g^2_{2}+\frac{13}{15}g^2_{1}
-6y_t^{2}-3\kappa_{10}^2\right) \right.     \nonumber \\
&-&\frac{1}{(4 \pi)^2}\left(\frac{3913}{450}g^4_{1}+\frac{33}{2}g^4_{2}+\frac{128}{9}g^4_{3}+g^2_{1}g^2_{2}
+\frac{136}{45}g^2_{1}g^2_{3}+8g^2_{2}g^2_{3}+
\left(\frac{2}{5}g^2_{1}+6g^2_{2} \right)y^2_{t}\right.   \nonumber \\
&+& \left. \left.  \left(\frac{4}{5}g^2_{1}+16g^2_{3} \right) \left( y^2_{t}+\kappa ^2_{10}\right) -
9\left( y^4_{t}+\kappa ^4_{10}\right)-9 y^2_{t}\left( y^2_{t}+\kappa ^2_{10}\right)-4 y^4_{t} \right) \right).
\end{eqnarray}%

The additional vectorlike matter fields also modify the RGE's
for the MSSM gauge couplings and the corresponding beta functions
can be found in \cite{Hempfling:1995rb}.                                                                                                                                                                                                                                                                                                                                                                                                                                                                                                                                                                                                                                                                                                                                                                                                                                                                                                                                                                                                                                                                                                                                                                                                                                                                                                                                                                     

\begin{figure}[t]
\centering \includegraphics[angle=0, width=13cm]{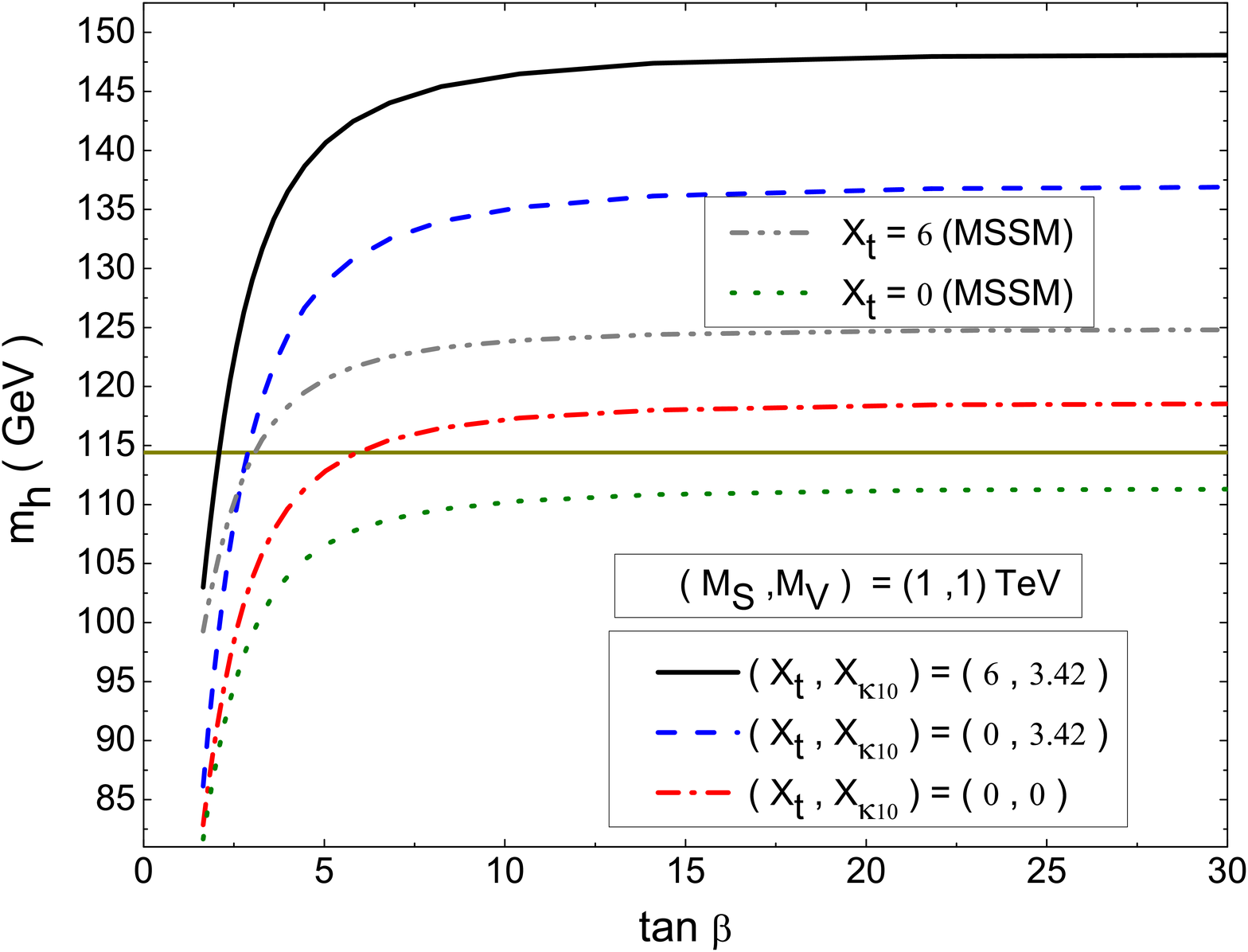}
\vspace{-1cm} \caption{ Upper bounds for the lightest CP-even Higgs boson mass vs
$\tan\beta$ for different maximal and minimal values of $X_t$,
$X_{\kappa_{10}}$, with $M_S=1$ TeV, $M_V=1$ TeV and $M_t=172.6$ GeV.
Dotted  line corresponds to MSSM ($X_t=0$).
Dashed-double dotted line describes MSSM with ($X_t=6$).
Dashed-dotted curve is for MSSM $+10+\overline{10}$.
$\kappa_{10} \approx 2$ at $M_{G}$. Dashed line shows Higgs mass with
$X_{\kappa_{10}}=2.95$ and $X_t=0$. Solid line corresponds to $X_{\kappa_{10}}=2.95$ and
 $X_t=6$.
} \label{mh102}
\end{figure}

\begin{figure}[t]
\centering \includegraphics[angle=0, width=13cm]{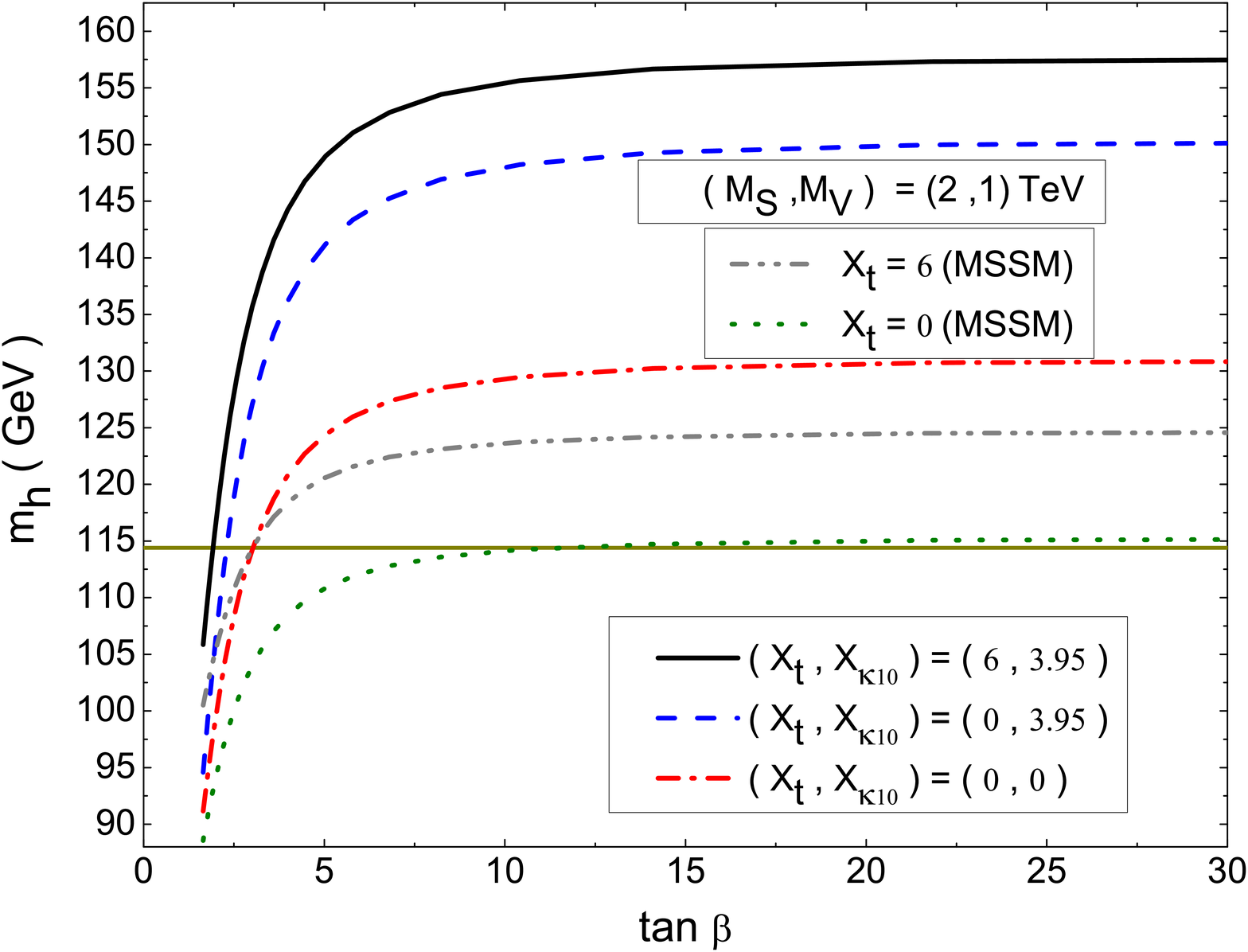}
\vspace{-1cm} \caption{Upper bounds for the lightest CP-even Higgs boson mass vs
$\tan\beta$ for different maximal and minimal values of $X_t$,
$X_{\kappa_{10}}$, with $M_S=2$ TeV, $M_V=1$ TeV and $M_t=172.6$ GeV.
Dotted  line corresponds to MSSM ($X_t=0$).
Dashed-double dotted line describes MSSM with ($X_t=6$).
Dashed-dotted curve is for MSSM $+10+\overline{10}$.
$\kappa_{10} \approx 2$ at $M_{G}$.
  Dashed line shows Higgs mass
 with $X_{\kappa_{10}}=3.95$ and $X_t=0$. Solid line corresponds to $X_{\kappa_{10}}=3.95$ and $X_t=6$.} \label{mh103}
\end{figure}

In our calculation, the weak scale ($M_Z$) value of the gauge and
top Yukawa couplings are evolved to the scale $M_G$ via the RGE's in
$\overline{DR}$ regularization scheme, where the scale $M_G$ is
defined to be one where $g_2=g_1$. We do not enforce an exact
unification of the strong coupling $g_3=g_2=g_1$ at scale $M_G$, since
a few percent deviation from the unification condition can be
assigned to unknown GUT scale threshold corrections. At the
scale $M_G$ we impose $\kappa_{10} \approx 2$, in order to obtain the maximal
value for $\kappa_{10}$ at low scale, which is consistent with the $T$
parameter constraints. In this case we can generate, according to Eq.
(\ref{max_10}), the maximal plausible values for Higgs mass. Our
goal is to achieve the maximal value for the CP-even Higgs mass and,
as we show  in Eq. (\ref{e1}), the Higgs mass is proportional  to
$\kappa_{10}^4$. The coupling $\kappa_{10}$, along with the gauge and top
Yukawa couplings, are evolved back to $M_Z$. In the evolution of
couplings, for the SUSY threshold correction we follow the effective
SUSY scale approach, according to which all SUSY particles are
assumed to lie at an effective scale \cite{Lang}. Below $M_{SUSY}$
we employ the non-SUSY RGEs. All of the couplings are iteratively
run between $M_Z$ and $M_G$ using two loop RGEs for both the
Yukawa and gauge couplings until a stable solution is obtained. Note
that $M_{S}$ and $M_{SUSY}$ are distinct parameters. As pointed out
in ref. \cite{Lang}, one can have a different set of values for stop
squark masses for a given effective $M_{SUSY}$ and so
correspondingly one considers different values of $M_{S}$ for
$M_{SUSY} = 200$ GeV.

Requiring $\delta T<0.1$ with $10+\overline{10}$ masses at $M_{V}=1$
TeV, we find that $\kappa _{10}(M_{V})<1.142$ at $(M_{V})$ scale
using the formula from ref. \cite{Lavoura:1992np}. The corresponding
$\kappa_{10}$ at GUT scale in this case is   $\kappa
_{10}(M_{G})\approx 2$. We find that the $S$--parameter constraint 
is automatically satisfied once the $T$--parameter constraint is met.

We can see from Eqs. (\ref{eq1}) -- (\ref{A1}) that to maximize the
CP-even Higgs boson mass we should not only take the maximal allowed
value for $\kappa_{10}$, we also need to have the maximal values for
the parameters $X_t$ and $X_{\kappa_{10}}$. According to Eq. (\ref{f1})
we find that $X_{\kappa_{10}}=2.95$, with $M_S=500$ GeV and $M_V=1$ TeV.
The value for $X_{\kappa_{10}}$ increases ( $X_{\kappa_{10}}=3.42$ ) if we
consider $M_S= 1$ TeV and $M_V=1$ TeV, while $X_{\kappa_{10}}=3.95$ for
$M_S= 2$ TeV and $M_V=1$ TeV.

We find that $M_V=$1 TeV is somehow the optimum value for the
vectorlike particle mass, especially because the $T$ parameter
constraint almost disappears for this value of $M_V$. On the other
hand Eq. (\ref{mm7}) does not allow very low values for $M_S$ if
significant corrections are to be realized. This is the reason why
we choose $M_S=0.5,\, \, 1$ and 2 TeV for our analysis.

In Figure \ref{mh101} we present the upper bounds for the CP-even Higgs boson
mass vs $\tan\beta$ with different maximal or minimal values of
$X_t$, $X_{\kappa_{10}}$ when $M_S=500$ GeV and $M_V=1$  TeV, and we
compare to the MSSM case. We take at scale $M_{G}$,
 $\kappa_{10} \approx 2$ to obtain the maximal effect for the lightest Higgs boson
 mass. As we see from Figure \ref{mh101}, for this choice of parameters the
 maximal values for Higgs mass is 141 GeV. In Figure \ref{mh102} we present
 the results for the case in which the mass for vectorlike matter is $M_V=1$ TeV and
 $M_S=1$ TeV too. In this case the CP-even higgs mass can be as large as
 148 GeV. Finally in Figure \ref{mh103} we consider $M_V=1$ TeV and
 $M_S=2$ TeV case and obtain the maximal value of  158 GeV for the Higgs mass.

\vspace{0.7cm}

\subsection{{MSSM }$\mathbf{+}$ $5$ $\mathbf{+}$ $\overline{5}$}

\begin{figure}[t]
\centering \includegraphics[angle=0, width=13cm]{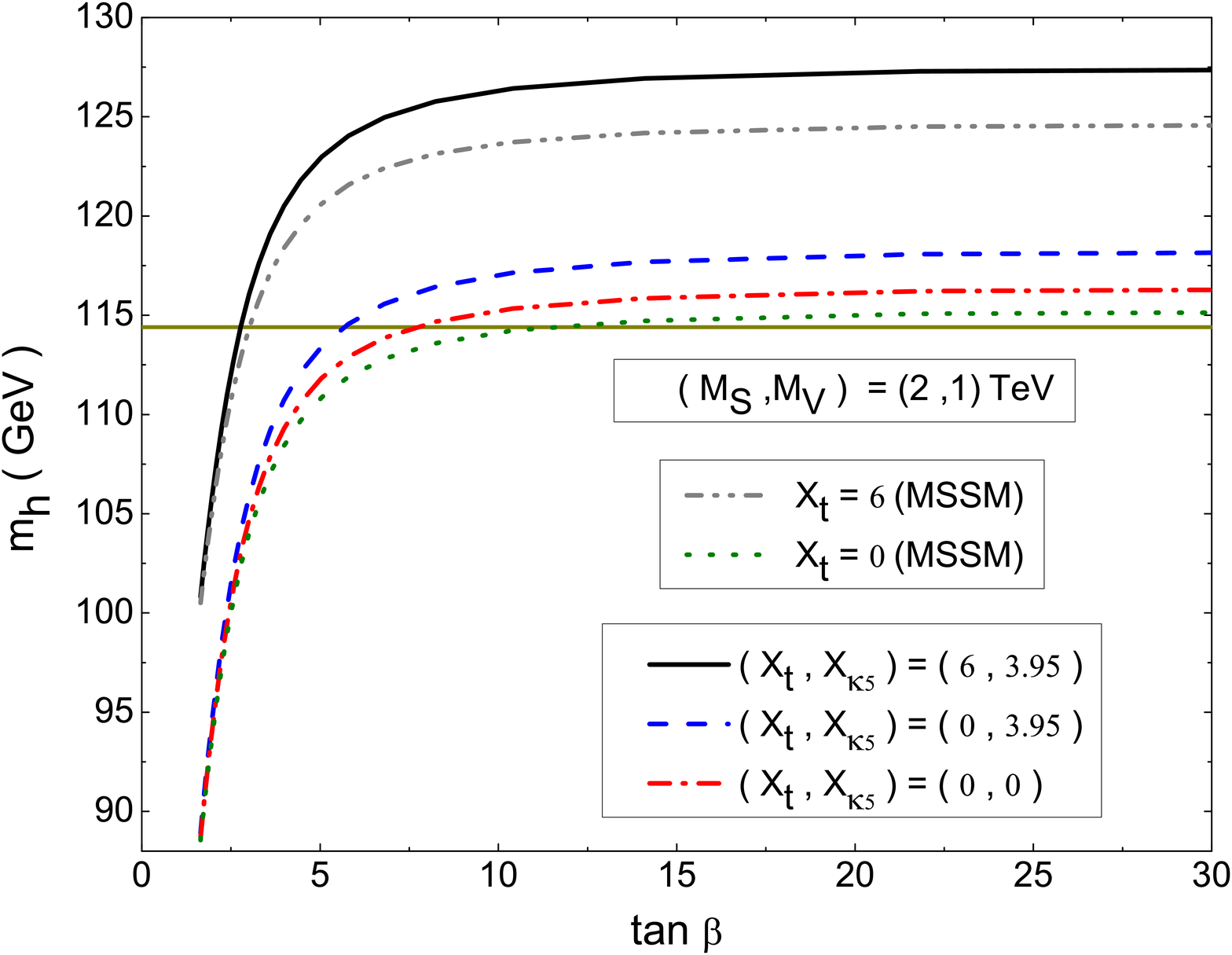}
\vspace{-1cm} \caption{Upper bounds for the lightest CP-even Higgs boson mass vs
$\tan\beta$ for different maximal and minimal values of $X_t$,
$X_{\kappa_{5}}$, with $M_S=2$ TeV, $M_V=1$ TeV and $M_t=172.6$ GeV.
Dotted  line corresponds to MSSM with ($X_t=0$).
Dashed-double dotted line describes MSSM with ($X_t=6$).
Dashed-dotted curve is for MSSM $+5+\overline{5}$.
$\kappa_{5} \approx 2$ at $M_{G}$.
  Dashed line shows Higgs mass
 with $X_{\kappa_{5}}=3.95$ and $X_t=0$. Solid line corresponds 
to $X_{\kappa_{5}}=3.95$ and $X_t=6$.} \label{mh51}
\end{figure}

In this subsection  we consider the case in which at the TeV
scale we have extra
 matter which belongs to the $5$ dimensional representation of
 $SU(5)$. This decomposes under the MSSM gauge symmetry as
follows:

\begin{equation}
5 + \overline{5}=L_{5}\left( 1,2,\frac{1}{2}\right) +%
\overline{L}_{5}\left( 1,2,-\frac{1}{2}\right) +D_{5}\left( \overline{3},1,%
\frac{1}{3}\right) +\overline{D}_{5}\left( 3,1,-\frac{1}{3}\right).
\end{equation}%

Our goal is to  generate new trilinear couplings of this extra
matter with the MSSM Higgs fields. The $5+\overline{5}$ itself
cannot generate this kind of  coupling. However, if we introduce an
MSSM singlet $S$, then Yukawa couplings of the form (in
$SU(5)$ notation) $\overline 5 \cdot S \cdot H_u$ and $5 \cdot S \cdot
H_d $ are permitted. In this case the MSSM superpotential has the
following additional contribution
\begin{equation}
W=\kappa _{5}L_{5}{S}H_{u}+\kappa _{5}^{\prime }\overline{L}%
_{5}S H_{d}+M_{V}\left( S \bar S + \overline{L}_{5}L_{5}+\overline{D_{5}}%
D_{5}\right).\label{dd66}
\end{equation}%
We take $\kappa _{5}\gg \kappa _{5}^{\prime }$, for the same
reason mentioned in the previous section. We also assume that there is an
additional symmetry which forbids the mixing of the vectorlike
particle with the MSSM matter fields. With this assumption the
singlet field $S$ cannot be identified with the right handed
neutrino.

Using the effective potential approach we calculate the additional
contribution to the CP even Higgs mass at one loop level. [A similar
calculation was done in ref. \cite{ b3}.]
\begin{eqnarray}
\left[ m_{h}^{2}\right] _{5} &=&-M_{Z}^{2}\cos ^{2}2\beta \left(
\frac{1}{8\pi ^{2}}\kappa_{5} ^{2}t_{V}\right)
+\frac{1}{4\pi ^{2}}\kappa_{5} ^{4}v^{2}\sin ^{2}\beta \left[ t_{V}+\frac{1%
}{2}X_{\kappa_{5} }\right],  \label{e3}
\end{eqnarray}%
 where we have assumed $M_{V}\gg M_{D}$ and
\begin{equation}
X_{\kappa_{5} }=\frac{4\widetilde{A}_{\kappa_{5} }^{2}\left(
3M_{S}^{2}+2M_{V}^{2}\right) -\widetilde{A}_{\kappa_{5}
}^{4}-8M_{S}^{2}M_{V}^{2}-10M_{S}^{4}}{6\left(
M_{S}^{2}+M_{V}^{2}\right) ^{2}} \label{X3}
\end{equation}%
and
\begin{equation}
t_{V}=\log \left( \frac{%
M_{S}^{2}+M_{V}^{2}}{M_{V}^{2}}\right). \\
\end{equation}
Here $\widetilde{A}_{\kappa_{5}}=A_{\kappa_{5}}-\mu \cot \beta $,
$A_{\kappa_{5}}$ is the $L_{5}-S$ soft mixing parameter and
$\mu$ is the MSSM Higgs bilinear mixing term.

 The RGE for $\kappa_5$ has the following form
\begin{eqnarray}
 \frac{d \kappa_{5}}{dt}&=&\frac{\kappa_{5}}{2(4 \pi)^2}  \left( \left(
3 g^2_{2}+\frac{3}{5}g^2_{1}
-4\kappa_{5}^{2}-3y_{t}^2\right) \right.     \nonumber \\
&-&\frac{1}{(4 \pi)^2}\left(\frac{237}{50}g^4_{1}+\frac{21}{2}g^4_{2}
+\frac{9}{5}g^2_{1}g^2_{2}+
\left(\frac{6}{5}g^2_{1}+6g^2_{2} \right)\kappa_{5}^2\right.   \nonumber \\
&+& \left. \left.  \left(\frac{4}{5}g^2_{1}+16g^2_{3} \right) y^2_{t} -
3\left( 3y^4_{t}+\kappa ^4_{5}\right)-3 \kappa_{5}^2\left( 3y^2_{t}+\kappa_{5}^2\right)-4 \kappa_{5}^4 \right) \right).
 \label{k5}
\end{eqnarray}%
Because the new
matter fields couple to $H_u$ (see Eq. (\ref{dd66})), there are
additional contribution to the RGE for $y_t$ which to two--loop level is given by
\begin{eqnarray}
 \frac{d y_t}{dt}&=&\frac{y_t}{2(4 \pi)^2}  \left( \left(
\frac{16}{3}g^2_{3}+3 g^2_{2}+\frac{13}{15}g^2_{1}
-6y_t^{2}-\kappa_{5}^2\right) \right.     \nonumber \\
&-&\frac{1}{(4 \pi)^2}\left(\frac{3133}{450}g^4_{1}+\frac{21}{2}g^4_{2}+\frac{32}{9}g^4_{3}+g^2_{1}g^2_{2}
+\frac{136}{45}g^2_{1}g^2_{3}+8g^2_{2}g^2_{3}+
\left(\frac{2}{5}g^2_{1}+6g^2_{2} \right)y^2_{t}\right.   \nonumber \\
&+& \left. \left.  \left(\frac{4}{5}g^2_{1}+16g^2_{3} \right)y^2_{t} -
3\left(3 y^4_{t}+\kappa ^4_{5}\right)-3 y^2_{t}\left(3 y^2_{t}+\kappa ^2_{5}\right)-4 y^4_{t} \right) \right).
\end{eqnarray}%

We can see from Eq. (\ref{k5}) that $\kappa_5$ cannot be as large
at $M_Z$ scale as $\kappa_{10}$ was. The reason for this is that in
the RGE for $\kappa_5$, in contrast to the case for $\kappa_{10}$,
the strong gauge coupling does not participate at one loop level.
Because of this we find that $\kappa_5(M_Z)=0.74$ for $ M_{S}=$ 2 TeV
and $ M_{V}=$ 1 TeV.

In Figure \ref{mh51} we present the upper bounds for the CP-even Higgs boson
mass vs $\tan\beta$ with different maximal or minimal values of
$X_t$, $X_{\kappa_{5}}$ with $M_S=2$ TeV and $M_V=1$  TeV and
which we compare with the MSSM case. We take
 $\kappa_{5} \approx 2$ at scale $M_{G}$ as before.
 For this choice of parameters the
 maximal value for the Higgs mass is 127.5 GeV.

\subsection{{MSSM }$\mathbf{+}$ $5$ $\mathbf{+}$ $\overline{5}$ $\mathbf{+}$ $10$ $\mathbf{+}$ $\overline{10}$}

In this section we will consider extra vectorlike matter belonging
to the representation $5+ \overline 5 + 10 + \overline{10}$  of
SU(5). There are two choices to consider here, namely with or
without two SM singlet fields. This does not affect the
perturbativity condition, but the presence of the singlets suggests
an underlying SO(10) gauge symmetry.

\vspace{0.3cm}
{\it Case I.} Without the singlet the MSSM superpotential acquires
the following additional contribution
\begin{eqnarray}
W&=&\kappa _{1}Q_{10}U_{10}H_{u}+ \kappa _{2}\overline{Q}_{10}
\overline{D}_{5}H_{u} + \kappa _{3}\overline{Q}_{10}
\overline{U}_{10}H_{d}+ \kappa _4{Q}_{10}
{D}_{5}H_{d}  \nonumber \\ &+&M_{V}\left( \overline{Q}_{10}Q_{10}+\overline{U}%
_{10}U_{10}+\overline{E}_{10}E_{10} +
\overline{L}_{5}L_{5}+\overline{D_{5}} D_{5}\right).\label{dd666}
\end{eqnarray}%
The new interaction yields the following additional contribution to
the MSSM CP-even Higgs boson mass
\begin{eqnarray}
\left[ m_{h}^{2}\right] _{1} =&-&M_{Z}^{2}\cos ^{2}2\beta \left(
\frac{3}{8\pi ^{2}}\kappa_{1} ^{2}t_{V}\right)
+\frac{3}{4\pi ^{2}}\kappa_{1} ^{4}v^{2}\sin ^{2}\beta \left[ t_{V}+\frac{1%
}{2}X_{\kappa_{1} }\right],\nonumber \\
&-&M_{Z}^{2}\cos ^{2}2\beta \left( \frac{3}{8\pi ^{2}}\kappa_{2}
^{2}t_{V}\right)
+\frac{3}{4\pi ^{2}}\kappa_{2} ^{4}v^{2}\sin ^{2}\beta \left[ t_{V}+\frac{1%
}{2}X_{\kappa_{2} }\right], \label{e1}
\end{eqnarray}%
where we have assumed $M_{V}\gg M_{D}$, and defined
\begin{equation}
X_{\kappa_{i} }=\frac{4\widetilde{A}_{\kappa_{i} }^{2}\left(
3M_{S}^{2}+2M_{V}^{2}\right) -\widetilde{A}_{\kappa_{i}
}^{4}-8M_{S}^{2}M_{V}^{2}-10M_{S}^{4}}{6\left(
M_{S}^{2}+M_{V}^{2}\right) ^{2}}, \label{X1}
\end{equation}%
and
\begin{equation}
t_{V}=\log \left( \frac{%
M_{S}^{2}+M_{V}^{2}}{M_{V}^{2}}\right), \label{mm7}
\end{equation}
 where $i=$ 1, 2, $\widetilde{A}_{\kappa_{i}}=A_{\kappa_{i}}-\mu \cot \beta $ 
and $A_{\kappa_{i}}$ is the soft mixing parameter.

In this case the lightest CP-even Higgs mass is
\begin{eqnarray}
m_h^2= [m^2_h]_{MSSM}+[m_h^2]_1
\end{eqnarray}
where the expression for $[m^2_h]_{MSSM}$ is given in Eq.
(\ref{eq2})

The RGEs for $\kappa_1$ and $\kappa_2$ are given to one--loop order by
\begin{eqnarray}
\frac{d \kappa_{1}}{dt} =\frac{\kappa_{1}}{2(4 \pi)^2}\left(
\frac{16}{3}g^2_{3}+3 g^2_{2}+\frac{13}{15}g^2_{1}
-6\kappa_{1}^{2}- 3\kappa_{2}^{2}-3y_{t}^2\right), \nonumber \\
\frac{d \kappa_{2}}{dt} =\frac{\kappa_{2}}{2(4 \pi)^2}\left(
\frac{16}{3}g^2_{3}+3 g^2_{2}+\frac{7}{15}g^2_{1}
-6\kappa_{2}^{2}-3\kappa_{2}^{2}-3y_{t}^2\right).
\end{eqnarray}%

Because the new matter couples to  $H_u$ (see Eq. (\ref{dd666})) there
is an additional contribution to the RGE for $y_t$ at one--loop
level:
\begin{eqnarray}
 \frac{d y_t}{dt}= \left[  \frac{d y_t}{dt}\right]
 _{MSSM}-\frac{3}{2(4 \pi)^2}\, y_t \kappa_{1}^2 -\frac{3}{2(4 \pi)^2}\, y_t \kappa_{2}^2.
\end{eqnarray}%

\begin{figure}[t]
\centering \includegraphics[angle=0, width=13cm]{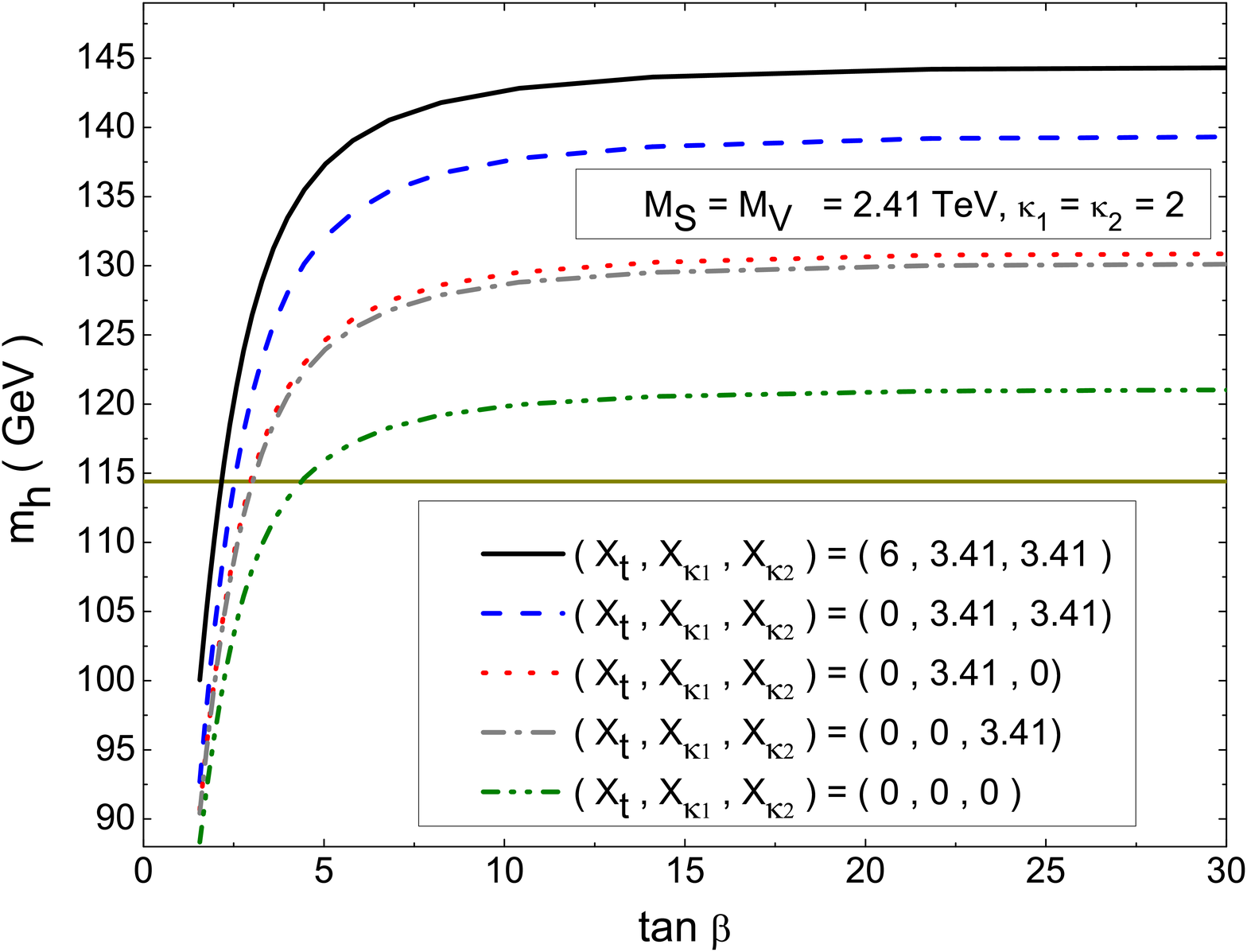}
\vspace{-1cm} \caption{Upper bounds for the lightest CP-even Higgs boson mass vs
$\tan\beta$ for different maximal and minimal values of $X_t$,
$X_{\kappa_{1}}$, $X_{\kappa_{2}}$, with $M_S=M_V=2.41$ TeV,  and
$M_t=172.6$ GeV. Dashed-double dotted curve describes the
logarithmical correction in the model. Dotted curve corresponds to
 the maximum value of $\kappa_1$ or $\kappa_2$.
Dashed curve corresponds to the maximum values of $\kappa_1$ and $\kappa_2$.
 Solid curve corresponds to the case when all corrections 
are taken to be maximum. } \label{mhtb105}
\end{figure}
\vspace{0.7cm}

In Figure \ref{mhtb105} we present the upper bounds for the CP-even Higgs boson
mass vs $\tan\beta$ for different maximal or minimal values of
$X_t$, $X_{\kappa_{1}}$, $X_{\kappa_{2}}$, with $M_S=M_V=2.41$ TeV,
and compare it with the MSSM case. We take
 $\kappa_i \approx 2$ at $M_{G}$ as before.
 For the given choice of parameters the
 maximal value of the Higgs mass is 144.5 GeV.

\vspace{0.3cm}
{\it Case II.} Next we consider the case when at low scale we have
vectorlike particles in $(16+ \overline{16})$ dimensional
representation of SO(10). The MSSM superpotential for this case
acquires the following additional contribution:
\begin{eqnarray}
W&=&\kappa _{1}Q_{10}U_{10}H_{u}+ \kappa _{2}\overline{Q}_{10}
\overline{D}_{5}H_{u} +   \kappa_3 L_5  S H_u  + \kappa _4{Q}_{10}
{D}_{5}H_{d} + \kappa
_{5}\overline{Q}_{10} \overline{U}_{10}H_{d} + \kappa_6 \overline{L_5} H_d \overline{S} \nonumber \\
 &+&M_{V}\left( \overline{Q}_{10}Q_{10}+\overline{U}%
_{10}U_{10}+\overline{E}_{10}E_{10} +
\overline{L}_{5}L_{5}+\overline{D_{5}} D_{5} + \overline{S} S
\right).\label{dd88}
\end{eqnarray}%

\begin{figure}[t]
\centering \includegraphics[angle=0, width=13cm]{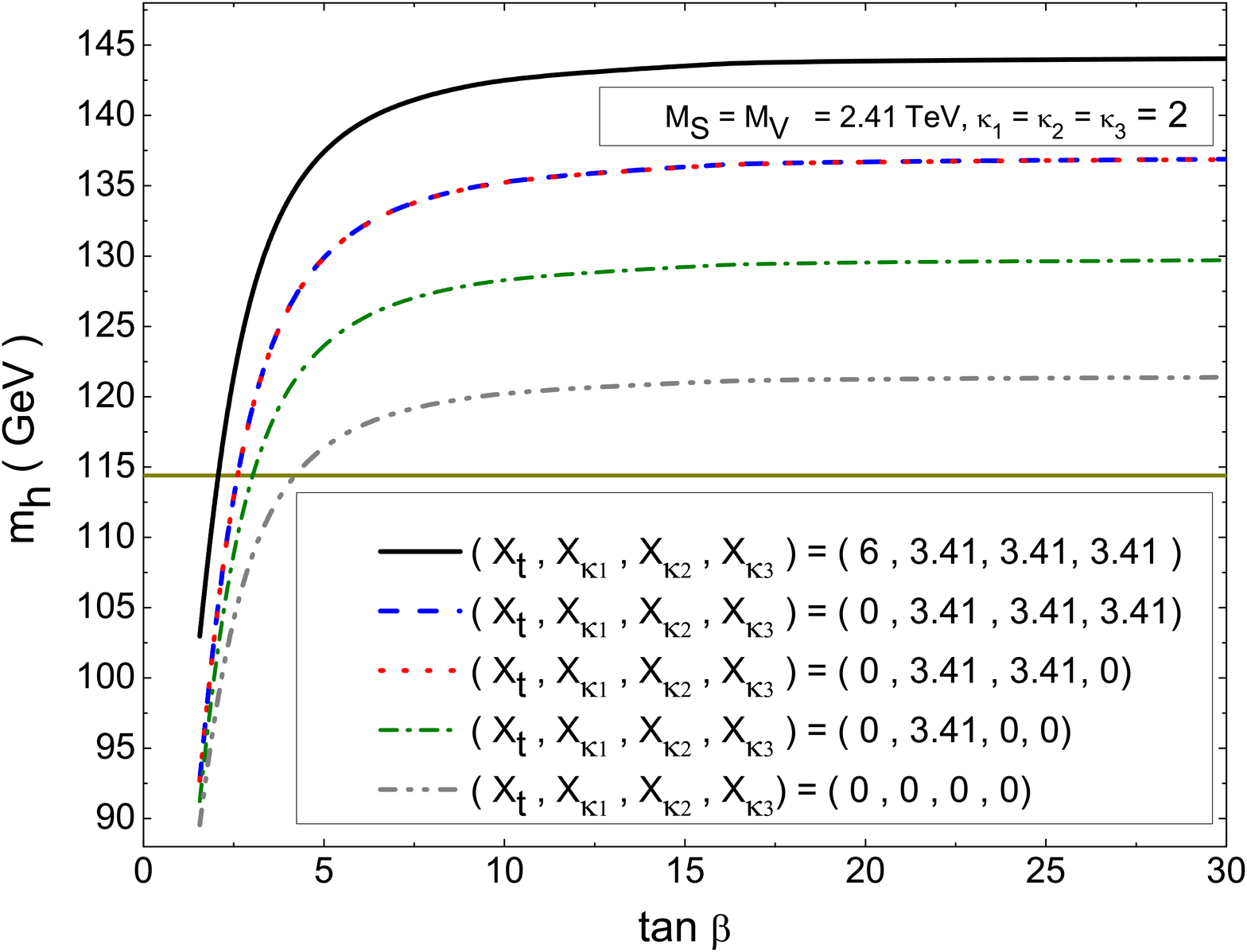}
\vspace{-1cm} \caption{Upper bounds for the lightest CP-even Higgs boson mass vs
$\tan\beta$ for different maximal and minimal values of $X_t$,
$X_{\kappa_{1}}$, $X_{\kappa_{2}}, X_{\kappa_{3}}$,   with
$M_S=M_V=2.41$ TeV and $M_t=172.6$ GeV. } \label{mhtb16}
\end{figure}

The new interactions provide the following additional contribution
to the MSSM CP-even Higgs boson mass
\begin{eqnarray}
\left[ m_{h}^{2}\right] _{2} =&-&M_{Z}^{2}\cos ^{2}2\beta \left(
\frac{3}{8\pi ^{2}}\kappa_{1} ^{2}t_{V}\right) +\frac{3}{4\pi
^{2}}\kappa_{1} ^{4}v^{2}\sin ^{2}\beta \left[ t_{V}+\frac{1
}{2}X_{\kappa_{1} }\right],\nonumber \\
&-&M_{Z}^{2}\cos ^{2}2\beta \left( \frac{3}{8\pi ^{2}}\kappa_{2}
^{2}t_{V}\right) +\frac{3}{4\pi ^{2}}\kappa_{2} ^{4}v^{2}\sin
^{2}\beta \left[ t_{V}+\frac{1
}{2}X_{\kappa_{2} }\right], \nonumber \\
 &-& M_{Z}^{2}\cos ^{2}2\beta \left(
\frac{1}{8\pi ^{2}}\kappa_{3} ^{2}t_{V}\right) +\frac{1}{4\pi
^{2}}\kappa_{3} ^{4}v^{2}\sin ^{2}\beta \left[ t_{V}+\frac{1
}{2}X_{\kappa_{3} }\right],
 \label{e1}
\end{eqnarray}%
 where we have assumed $M_{V}\gg M_{D}$, and defined
\begin{equation}
X_{\kappa_{i} }=\frac{4\widetilde{A}_{\kappa_{i} }^{2}\left(
3M_{S}^{2}+2M_{V}^{2}\right) -\widetilde{A}_{\kappa_{i}
}^{4}-8M_{S}^{2}M_{V}^{2}-10M_{S}^{4}}{6\left(
M_{S}^{2}+M_{V}^{2}\right) ^{2}}, \label{ww1}
\end{equation}%
and
\begin{equation}
t_{V}=\log \left( \frac{%
M_{S}^{2}+M_{V}^{2}}{M_{V}^{2}}\right). \label{mm8}
\end{equation}
Here $i=1,\, 2, \, 3$ and $\widetilde{A}_{\kappa_{i}}=A_{\kappa_{i}}-\mu \cot \beta $.
The lightest CP-even Higgs mass is
\begin{eqnarray}
m_h^2= [m^2_h]_{MSSM}+[m_h^2]_2,
\end{eqnarray}
where the expression for $[m^2_h]_{MSSM}$ is given in Eq.
(\ref{eq2}).

 The  RGEs for $\kappa_i$ are given by:
\begin{eqnarray}
\frac{d \kappa_{1}}{dt} &=&\frac{\kappa_{1}}{2(4 \pi)^2}\left(
\frac{16}{3}g^2_{3}+3 g^2_{2}+\frac{13}{15}g^2_{1}
-6\kappa_{1}^{2}- 3\kappa_{2}^{2}- \kappa_{3}^{2}-3y_{t}^2\right), \nonumber \\
\frac{d \kappa_{2}}{dt} &=&\frac{\kappa_{2}}{2(4 \pi)^2}\left(
\frac{16}{3}g^2_{3}+3 g^2_{2}+\frac{7}{15}g^2_{1}
-6\kappa_{2}^{2}-3\kappa_{2}^{2}-\kappa_{3}^{2}-3y_{t}^2\right), \nonumber \\
\frac{d \kappa_{3}}{dt} &=&\frac{\kappa_{3}}{2(4 \pi)^2}\left(
3g_2^2+\frac{3}{5}g_1^2 -4\kappa_{3}^{2}- 3\kappa_{1}^{2}
-3\kappa_{2}^{2}-3y_{t}^2\right).
\label{k3} 
\end{eqnarray}%

The RGE for $y_t$ is modified as follows:
\begin{eqnarray}
 \frac{d y_t}{dt}= \left[  \frac{d y_t}{dt}\right]
 _{MSSM}-\frac{3}{2(4 \pi)^2}\, y_t \kappa_{1}^2 -\frac{3}{2(4 \pi)^2}\, y_t \kappa_{2}^2 -\frac{1}{2(4 \pi)^2}\, y_t \kappa_{3}^2.
\end{eqnarray}%

In Figure \ref{mhtb16} we present the upper bounds for the CP-even Higgs boson
mass vs $\tan\beta$ for different maximal minimal values of
$X_t$, $X_{\kappa_{1}}$, $X_{\kappa_{2}}$, $X_{\kappa_{3}}$, with
$M_S=M_V=2.41$ TeV, and compare it with the MSSM case. We take
 $\kappa_i \approx 2$ at $M_{G}$ as before.
 For the given choice of parameters the
 maximal value of the Higgs mass is 143.9 GeV.
It is worth noting that the resultant Higgs mass bound
for {\it Case II} more or less coincides with what we found for
{\it Case I} (see Figure \ref{mhtb105}). This stems from the fact
that the contribution at `low' scale from the new coupling
$\kappa_{3}$ is small due to the absence of the strong coupling (see Eq. (\ref{k3})).

\section{Little Hierarchy problem}

\subsection{MSSM}

\begin{figure}[th]
\centering \includegraphics[angle=0, width=13cm]{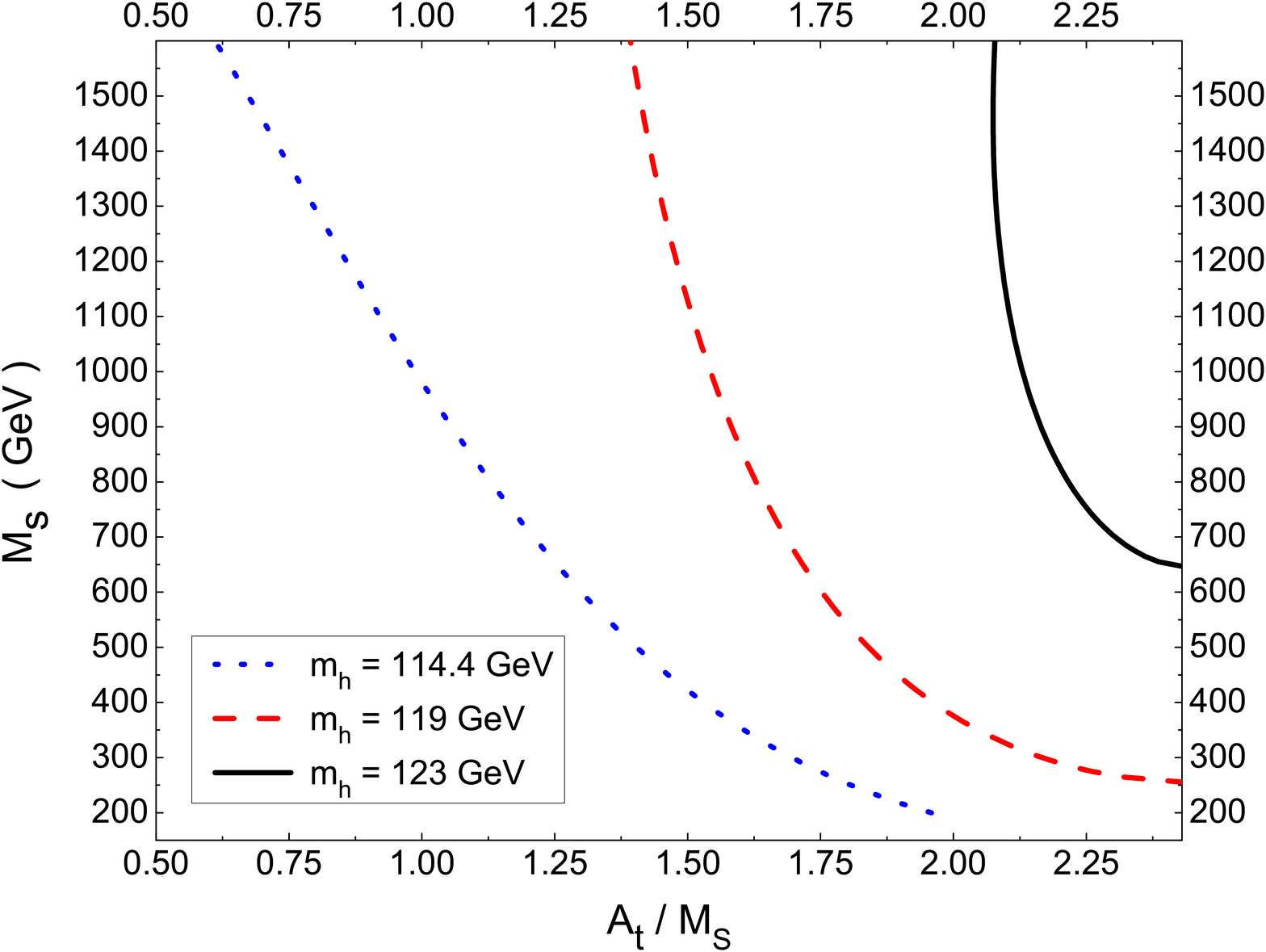}
\vspace{-1cm} \caption{ {$M_S$ versus  $A_t/M_S$ for
different values of CP-even Higgs mass in MSSM. Dotted, dashed
and solid line  corresponds to Higgs mass $m_h=$114.4, 119  and 123 GeV
respectively and  $\tan\beta=10$.} } \label{HB}
\end{figure}

As discussed in Section 3,   in the MSSM at tree level the lightest
CP even Higgs boson  mass is bounded from above by the mass of the $Z$
boson
\begin{equation}
m^2_h < M^2_Z\cos 2\beta.
\end{equation}
It requires large radiative corrections in order to push the lightest
Higgs mass above the LEP2 limit. We can see that there are two kind
of correction (see Eq. (7)), one proportional to
$m_{t}^4\log(M^2_S/m_{t}^2)$, where $M_{S} =\sqrt{m_{%
\widetilde{t}_{1}}m_{\widetilde{t}_{2}}}$, and the second
proportional to $A_t$.  As we can see from Figure \ref{HB}, if the Higgs
mass turns out to be much heavier then 114.4 GeV,  we need not only
a large trilinear soft SUSY breaking $A_t$ term but also heavy stop
quark masses.

On the other hand, the mass of the $Z$ boson ($M_Z \simeq 91$ GeV) is
given from the minimization of the scalar potential as (for
$\tan\beta \gtrsim 5$)
\begin{equation}
\frac{1}{2}M_{Z}^{2}\simeq -\mu ^{2}-m_{H_{u}}^{2},  \label{e4}
\end{equation}
and the radiative correction to the soft scalar mass
squared for $H_u$ is proportional to top squark masses
\begin{equation}
\Delta m_{H_{u}}^{2}(M_{Z})=-\frac{3y_{t}^{2}(M_{Z})}{4\pi
^{2}}M_{S}^{2}\ln \frac{\Lambda }{M_{S}}, \label{fff1}
\end{equation}%
where $\Lambda$ is a more fundamental
scale, such as $M_G$.

Thus, in the MSSM  one needs to have heavy top
squarks to generate the lightest Higgs mass above the LEP2 bound, while on the
other hand from Eqs. (\ref{e4}) and (\ref{fff1}) we see that some
fine tuning is needed to get the correct $Z$ boson mass. This is known
as the {\it little hierarchy problem}. To see how much fine tuning is
needed to satisfy Eq. (\ref{e4}) we  performed semi-analytic calculation
for the MSSM sparticle spectra with the following boundary
conditions
\begin{equation}
\{\alpha _{G},\,\, M_{G},\,\, y_{t}(M_{G})\}=\{{1}/{24.32},\,\,
2.0\times 10^{16},\,\, 0.512\}.
\end{equation}
We express the MSSM sparticle masses at the scale $M_{Z}$ in terms
of the GUT/Planck scale fundamental parameters ($m_0, m_{1/2},
A_{t_{0}}$) and the Higgs bi-linear mixing term $\mu$,
by integrating the one loop renormalization group equations
\cite{RGE}. For example, the gaugino masses at $M_{Z}$ scale are
\begin{equation}
\{M_{1}\left( M_{Z}\right) ,M_{2}\left( M_{Z}\right) ,M_{3}\left(
M_{Z}\right) \}=\{0.412,0.822,2.844\}m_{1/2}.
\end{equation}%
The scalar particle masses, $A_{t}$  and $\mu$ at the $M_{Z}$ scale
are given by
\begin{eqnarray}
-m_{H_{u}}^{2}\left( M_{Z}\right)
&=&2.72m_{1/2}^{2}+0.091m_{0}^{2}+0.1A_{t_{0}}^{2}-0.43 m_{1/2}A_{t_{0}} \\
m_{Q_{t}}^{2}\left( M_{Z}\right)
&=&5.71m_{1/2}^{2}+0.64m_{0}^{2}-0.033A_{t_{0}}^{2}+0.15m_{1/2}A_{t_{0}}\\
m_{U_{t}}^{2}\left( M_{Z}\right)
&=&4.2m_{1/2}^{2}+0.27m_{0}^{2}-0.07A_{t_{0}}^{2}+0.29
m_{1/2}A_{t_{0}}\\
A_{t}\left( M_{Z}\right) &=&-2.3m_{1/2}+0.27 A_{t_{0}}  \label{e6} \\
m_{Q_{1,2}}^{2}\left( M_{Z}\right) &=&6.79 m_{1/2}^{2}+m_{0}^{2} \\
m_{U_{1,2}}^{2}\left( M_{Z}\right) &=&6.37m_{1/2}^{2}+m_{0}^{2} \\
m_{D_{1,2,3}}^{2}\left( M_{Z}\right) &=&6.32m_{1/2}^{2}+m_{0}^{2} \\
m_{L_{1,2,3}}^{2}\left( M_{Z}\right) &=&0.52m_{1/2}^{2}+m_{0}^{2} \\
m_{E_{1,2,3}}^{2}\left( M_{Z}\right) &=&0.15m_{1/2}^{2}+m_{0}^{2}\\
\mu ^{2}\left( M_{Z}\right) &=&1.02\mu _{0}^{2}.
\end{eqnarray}%
where the subscript 1,2,3 are families indices and $\mu_0$ is the value of
$\mu$ at $M_G$.

Using Eq. (\ref{e4}) we can also express the dominant contribution to $Z$ 
boson mass in terms of fundamental parameters:
\begin{equation}
M_{Z}^{2} \simeq -2.04\mu ^{2}+5.44 m_{1/2}^{2}+0.183 m_{0}^{2}+0.2
A_{t_{0}}^{2}-0.87 m_{1/2}A_{t_{0}}.\label{mz1}
\end{equation}

\begin{figure}[t]
\centering \includegraphics[angle=0, width=13cm]{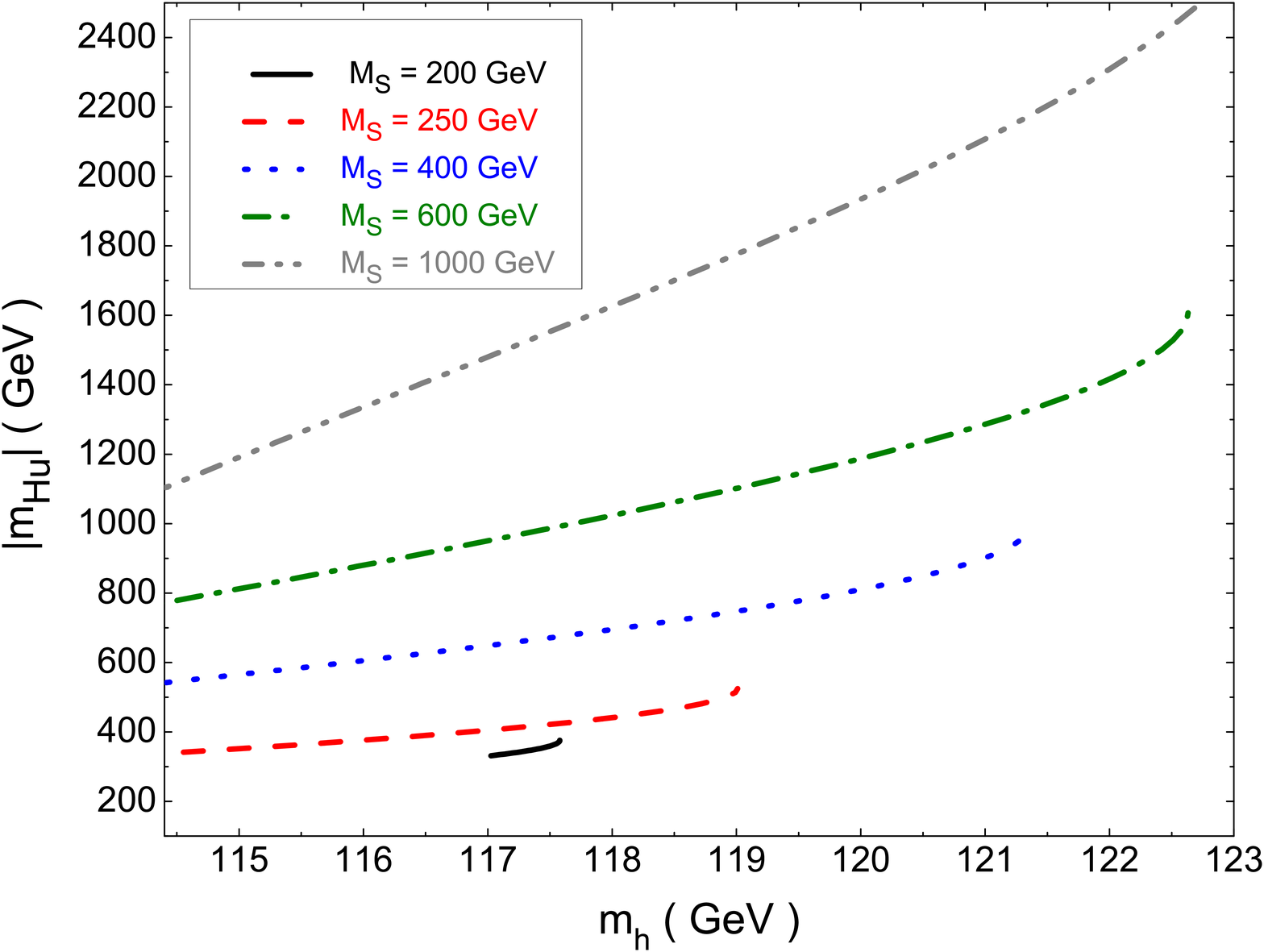}
\vspace{-1cm} \caption{$\left| m_{H_{U}}\right| $ versus
CP-even Higgs mass  in the MSSM for different values of $M_S$.
Solid, dashed, dotted, dashed-dotted and dashed-dotted-dotted curve corresponds $M_S =$
200 GeV, 250 GeV, 400 GeV, 600 GeV and 1000 GeV respectively.  In each  case
$A_t$ varies  in the interval  $0< \left| A_t/M_S \right| < \sqrt
{6}$. } \label{MSSM}
\end{figure}

The magnitude of $\left| m_{H_{U}}\right| $ shows how much
fine tuning is needed to satisfy the minimization condition in the MSSM,
(see Eq. (\ref{e4})). We present in Figure
\ref{MSSM} $\left| m_{H_{U}}\right| $ versus the CP-even Higgs mass for
different values of $M_S$. In each  case $A_t$ varies  in the
interval  $0< \left| A_t/M_S \right| < \sqrt {6}$. This is the
reason why we find different values for the Higgs masses for
different choice of $M_S$.

\begin{figure}[t]
\centering \includegraphics[angle=0, width=13cm]{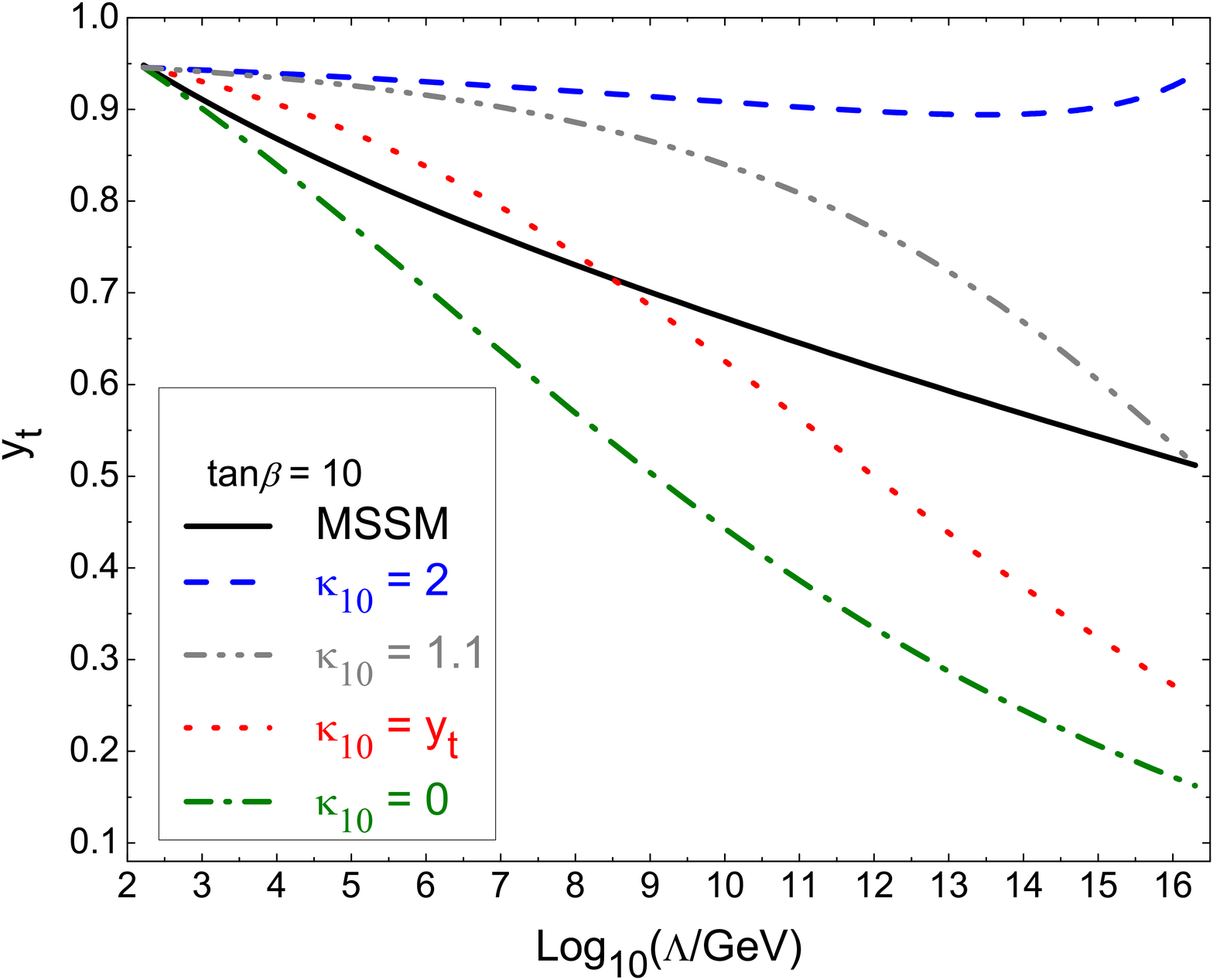}
\vspace{-1cm} \caption{Top Yukawa $y_t$
coupling versus  $\mbox{Log}_{10}( \Lambda /\mbox{GeV})$ for
$\tan\beta=10$. Dashed, dotted, dashed-dotted and dashed-dotted-dotted 
lines correspond to MSSM$+10+\bar{10}$ with $\kappa_{10}=2, \,y_t, \,0, \,1.1$ respectively.
Solid line correspond  to the MSSM case.} \label{yukawa}
\end{figure}

We find that the new Yukawa couplings of $H_u$ to the vectorlike matter,
which helps in raising $m_h$, also has the effect of raising the soft
Higgs mass parameter $m_{H_{u}}^2$, which tends to exacerbate the
little hierarchy problem. However,  when the new Yukawa couplings of
$H_u$ are relatively small, the little hierarchy problem improves
relative to the MSSM, since the cumulative effect of the top Yukawa
coupling $y_t$ on the running of $m_{H_{u}}^2$ becomes smaller than
in MSSM. This comes about since the value of $y_t$ is smaller at the
scale $M_G$ compared to MSSM  for certain values of the new Yukawa
coupling. This result is displayed in Figure \ref{yukawa} for the
MSSM$+10+\overline{10}$ case. There is also contribution from
radiative correction involving the gluon and gluino, since, as we show in Figure \ref{gu},
introducing new vectorlike matter at low scale slows the running of
strong coupling compared to the MSSM case. As a result
the cumulative effect of the strong interaction to the running of
colored particle masses is smaller than in  MSSM. These two effects, as
we show in the next two sections, enable us to improve the little hierarchy
problem compared to the MSSM.

\subsection{ MSSM$+10+\overline{10}$}

Next let us consider how the little hierarchy problem can be improved in
the MSSM$+10+\overline{10}$ case. We will consider two extreme values for
the coupling $\kappa_{10}$, namely $\kappa_{10}(M_{G})=2$ and
$\kappa_{10}(M_{G})=0$ to show how much little hierarchy has changed
for this case.

\begin{figure}[htbp]
\centering \includegraphics[angle=0, width=13cm]{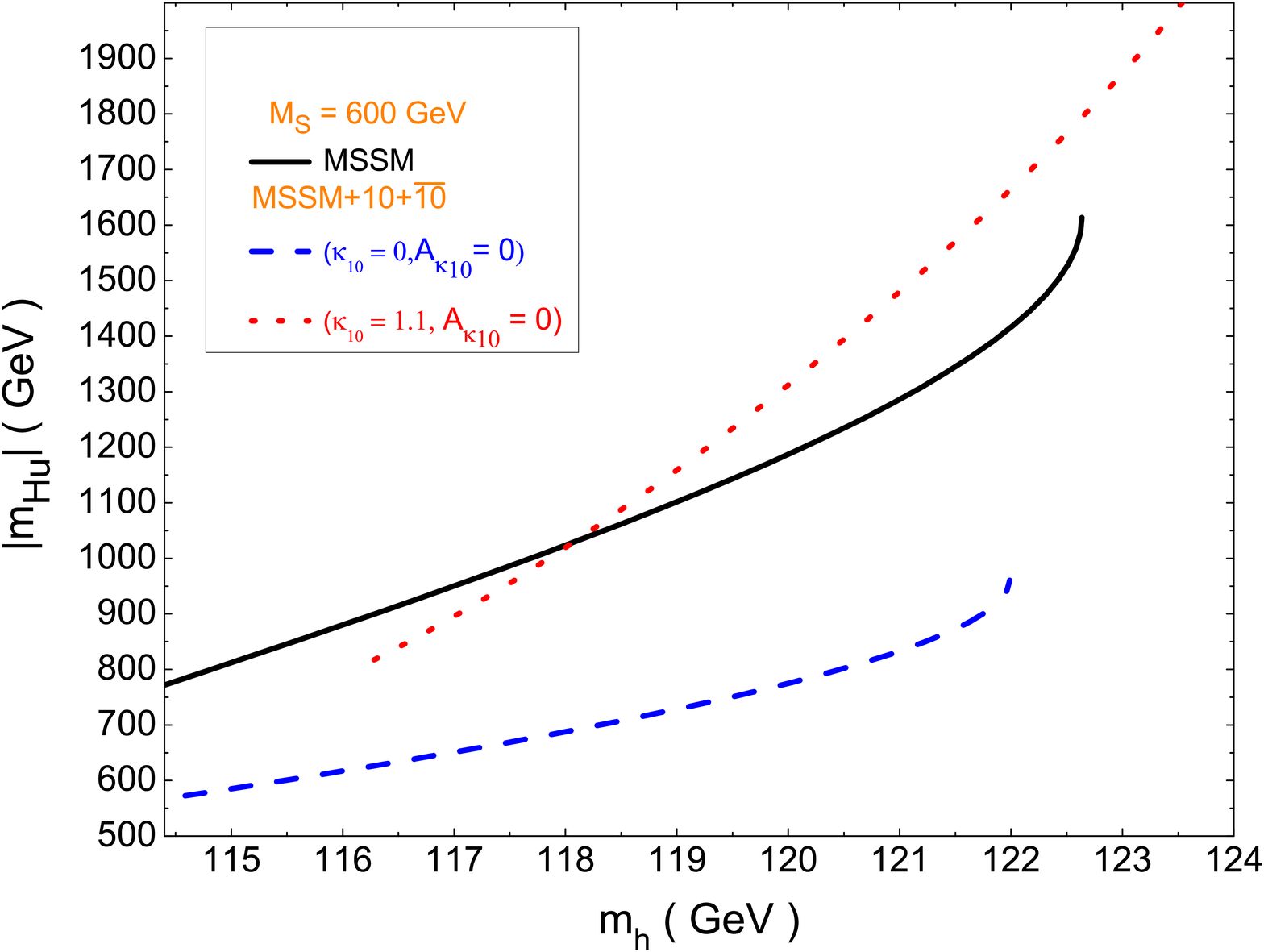}
\vspace{-1cm} \caption{$\left| m_{H_{U}}\right| $  versus
the CP even Higgs mass   $m_{h}$, with $M_S=600$ GeV. Solid line
correspond to the MSSM case. Dashed and dotted curve correspond
to MSSM$+ 10 + \overline{10}$, with $\kappa_{10}=1.1$ and $ \kappa_{10}=0$
at $M_G$. } \label{1010bar1}
\end{figure}

\begin{figure}[bt]
\centering \includegraphics[angle=0, width=13cm]{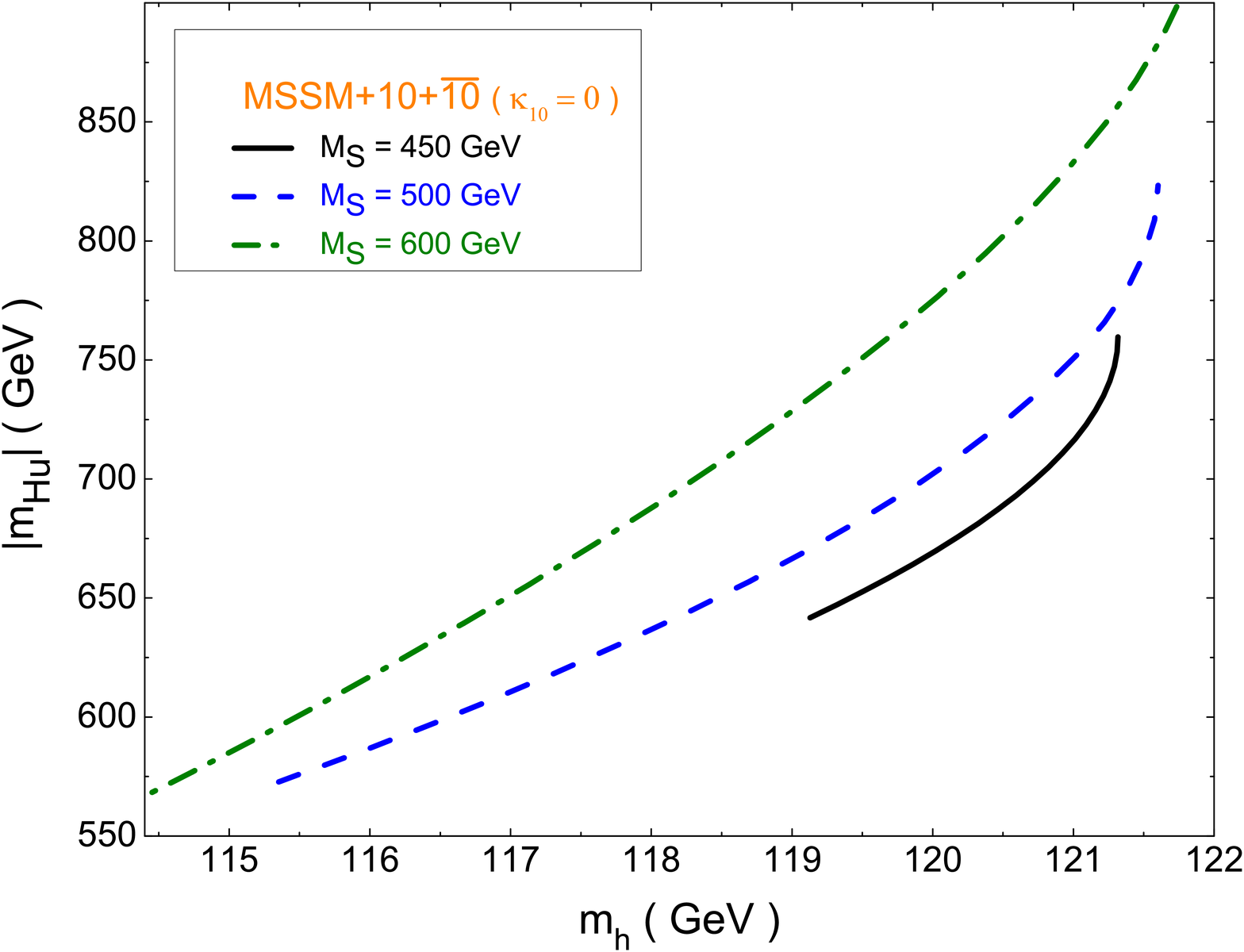}
\vspace{-1cm} \caption{$\left| m_{H_{U}}\right| $  versus
$m_{h}$ for MSSM$+10+ \overline{10}$, with ($A_{t}<0$). Solid, dashed 
and dashed-dotted curve correspond to $M_S=450, 500$ and 600 GeV 
respectively. For all cases $\kappa_{10}=0$ at $M_G$.} \label{1010bar2}
\end{figure}

\vspace{0.3cm}
{\it Case I.} Using the boundary conditions
\begin{equation}
\{\alpha _{G},\,\, M_{G},\,\, y_{t}(M_{G}),\,\, \kappa
_{10}(M_{G})\}=\{{1}/{8.55},\,\, 2.0\times 10^{16},\,\,0.94,\,\,
2\}.
\end{equation}
and RGEs from Appendix A, we obtain the sparticle spectrum.
For the gaugino masses,
\begin{equation}
\{M_{1}\left( M_{Z}\right) ,M_{2}\left( M_{Z}\right) ,M_{3}\left(
M_{Z}\right) \}=\{0.145,0.289,1\}m_{1/2},
\end{equation}%
while the MSSM scalar masses along with $\mu ^{2}$, $A_{t}$ and $A_{\kappa
_{10}}$ are given by
\begin{eqnarray}
-m_{H_{u}}^{2}\left( M_{Z}\right)
&=&3.85m_{1/2}^{2}+0.95m_{0}^{2}+0.04A_{t_{0}}^{2}+0.012A_{\kappa
_{10_{0}}}^{2}  \nonumber \\
&&-0.12 m_{1/2}A_{t_{0}}+0.06
m_{1/2}A_{\kappa_{10_{0}}}-0.043A_{t_{0}}A_{\kappa _{10_{0}}}\\
m_{Q_{t}}^{2}\left( M_{Z}\right)
&=&2.98m_{1/2}^{2}+0.73m_{0}^{2}-0.031A_{t_{0}}^{2}-0.002A_{\kappa_{10_{0}}}^{2}  \nonumber \\
&&+0.11m_{1/2}A_{t_{0}}-0.071m_{1/2}A_{\kappa_{10_{0}}}+0.026A_{t_{0}}A_{\kappa _{10_{0}}}\\
m_{U_{t}}^{2}\left( M_{Z}\right)
&=&2.04m_{1/2}^{2}+0.45m_{0}^{2}-0.062A_{t_{0}}^{2}-0.003A_{\kappa_{10_{0}}}^{2} \nonumber \\
&&+0.23m_{1/2}A_{t_{0}}-0.14m_{1/2}A_{\kappa_{10_{0}}}+0.052A_{t_{0}}A_{\kappa _{10_{0}}}
\end{eqnarray}
\begin{eqnarray}
A_{t}\left( M_{Z}\right) &=&-1.02m_{1/2}+0.2A_{t_{0}}-0.13A_{\kappa_{10_{0}}}  \label{e8} \\
A_{\kappa _{10}}\left( M_{Z}\right)
&=&-0.71m_{1/2}-0.093A_{t_{0}}+0.065A_{\kappa _{10_{0}}}  \label{e9} \\
m_{Q_{1,2}}^{2}\left( M_{Z}\right) &=&3.63 m_{1/2}^{2}+m_{0}^{2} \\
m_{U_{1,2}}^{2}\left( M_{Z}\right) &=&3.33m_{1/2}^{2}+m_{0}^{2} \\
\mu ^{2}\left( M_{Z}\right) &=&0.105\mu _{0}^{2} \\
m_{D_{1,2,3}}^{2}\left( M_{Z}\right) &=&m_{0}^{2}+3.29m_{1/2}^{2} \\
m_{L_{1,2,3}}^{2}\left( M_{Z}\right) &=&m_{0}^{2}+0.37m_{1/2}^{2} \\
m_{E_{1,2,3}}^{2}\left( M_{Z}\right) &=&m_{0}^{2}+0.122m_{1/2}^{2}.
\end{eqnarray}
The spectrum for the vectorlike matter is given as
\begin{eqnarray}
m_{Q_{10}}^{2}\left( M_{Z}\right)
&=&2.86m_{1/2}^{2}+0.62m_{0}^{2}+0.017A_{t_{0}}^{2}-0.003A_{\kappa_{10_{0}}}^{2} \nonumber \\
&&-0.073m_{1/2}A_{t_{0}}+0.051m_{1/2}A_{\kappa_{10_{0}}}-0.011A_{t_{0}}A_{\kappa
_{10_{0}}}\\
m_{U_{10}}^{2}\left( M_{Z}\right)
&=&1.81m_{1/2}^{2}+0.25m_{0}^{2}+0.035A_{t_{0}}^{2}-0.005A_{\kappa_{10_{0}}}^{2}\nonumber \\
&&-0.15m_{1/2}A_{t_{0}}+0.1m_{1/2}A_{\kappa_{10_{0}}}-0.023A_{t_{0}}A_{\kappa _{10_{0}}}\\
m_{E_{10}}^{2}\left( M_{Z}\right) &=&m_{0}^{2}+0.122m_{1/2}^{2}.
\end{eqnarray}%

For this case the dominant contribution of $Z$-boson mass has the following expression
\begin{eqnarray}
M_{Z}^{2} &\simeq &-0.21\mu
_{0}^{2}+7.7m_{1/2}^{2}+1.91m_{0}{}^{2}+0.08A_{t_{0}}^{2}+0.024A_{%
\kappa _{10_{0}}}^{2}  \nonumber \\
&&-0.24 m_{1/2}A_{t_{0}}+0.12m_{1/2}A_{\kappa _{10_{0}}}-0.094A_{t_{0}}A_{%
\kappa _{10_{0}}}. \label{mz2}
\end{eqnarray}
We see that the coefficient of $m_{1/2}^{2}$ in this
expression has increased as compared to MSSM case (see
Eq.(\ref{mz1})), and so we expect fine tuning to be worse in this case.
This  is  related to the value of $\kappa _{10}(M_{G})$. If we
reduce $\kappa _{10}(M_{G})$, $y_{t}(M_{G})$ is reduced (see Figure
(\ref{yukawa})), and as a result the coefficient of $m_{1/2}^{2}$ in
the $M_{Z}^{2}$ expression is reduced. We find that for some values
of $\kappa_{10}(M_{G})$ the top Yukawa coupling $y_{t}(M_{G})$ is smaller 
than $y_{t}^{MSSM}(M_{G})$, its value in the MSSM. This enables us to
reduce the degree of fine tuning for the case $MSSM+10+ \overline
{10}$ compared to the MSSM.

 However, with smaller values of $\kappa
_{10}(M_{G})$, the value for the Higgs mass $m_{h}$ will be
lower. Thus, we need to find an optimum value of $\kappa_{10}(M_{G})$,
between $\ 2$ and $0$, which gives a sufficiently large value for the Higgs mass,
but at the same time yields smaller value of $\left| m_{H_{u}}\right| $.

\vspace{0.3cm}
{\it Case II.} We next study the changes brought about by setting $\kappa
_{10}(M_{G})=0$ and by applying the following boundary conditions
\begin{equation}
\{\alpha _{G},M_{G},y_{t}(M_{G}),\kappa _{10}(M_{G}), A_{\kappa
_{10}}(M_G)\}=\{{1}/{8.55}%
,2.0\times 10^{16},0.163,0,0\}.
\end{equation}
For the MSSM spectrum and related quantities, we find
\begin{eqnarray}
-m_{H_{u}}^{2}\left( M_{Z}\right)
&=& 2.59 m_{1/2}^{2}-0.15m_{0}^{2}+0.12A_{t_{0}}^{2}-0.76 m_{1/2}A_{t_{0}}  \\
m_{Q_{t}}^{2}\left( M_{Z}\right)
&=&2.64m_{1/2}^{2}+0.72m_{0}^{2}-0.041A_{t_{0}}^{2}+0.25m_{1/2}A_{t_{0}} \\
m_{U_{t}}^{2}\left( M_{Z}\right)
&=&1.35m_{1/2}^{2}+0.44m_{0}^{2}-0.082A_{t_{0}}^{2}+0.51m_{1/2}A_{t_{0}} \\
A_{t}\left( M_{Z}\right)  &=&-2.14m_{1/2}+0.44A_{t_{0}} 
\end{eqnarray}
\begin{eqnarray}
A_{\kappa _{10}}\left( M_{Z}\right)
&=&-3.02m_{1/2}-0.28A_{t_{0}}+A_{\kappa _{10_{0}}} \\
m_{Q_{1,2}}^{2}\left( M_{Z}\right) &=&3.63 m_{1/2}^{2}+m_{0}^{2} \\
m_{U_{1,2}}^{2}\left( M_{Z}\right) &=&3.33m_{1/2}^{2}+m_{0}^{2} \\
\mu ^{2}\left( M_{Z}\right) &=&1.9\mu _{0}^{2} \\
m_{D_{1,2,3}}^{2}\left( M_{Z}\right) &=&m_{0}^{2}+3.29m_{1/2}^{2} \\
m_{L_{1,2,3}}^{2}\left( M_{Z}\right) &=&m_{0}^{2}+0.374m_{1/2}^{2} \\
m_{E_{1,2,3}}^{2}\left( M_{Z}\right) &=&m_{0}^{2}+0.122m_{1/2}^{2}\\
m_{Q_{10}}^{2}\left( M_{Z}\right)
&=&3.63m_{1/2}^{2}+m_{0}^{2}\\
m_{U_{10}}^{2}\left( M_{Z}\right)
&=&3.33m_{1/2}^{2}+m_{0}^{2} \\
m_{E_{10}}^{2}\left( M_{Z}\right) &=&m_{0}^{2}+0.122m_{1/2}^{2}.
\end{eqnarray}%
The dominant contribution of $Z$-boson mass has the following expression
\begin{eqnarray}
M_{Z}^{2} &\simeq &-3.78\mu_{0}^{2}+5.19m_{1/2}^{2}-0.31m_{0}^{2}+0.25A_{t_{0}}^{2}-1.52m_{1/2}A_{t_{0}}.
\end{eqnarray}
In Figure \ref{1010bar1} we plot  $\left| m_{H_{U}}\right| $ versus
the lightest CP even Higgs mass  $m_{h}$, with $M_S=600$ GeV. We compare
two cases, when $\kappa_{10}=1.1$ and $ \kappa_{10}=0$ at $M_G$. The
case $\kappa=1.1$ is interesting in the sense that in this case the
value of top Yukawa coupling is the same as in the MSSM. 
We can see in Figure \ref{1010bar1} that for a Higgs mass 
less than 118 GeV, the fine tuning responsible for little hierarchy problem is
less severe, while larger than these values the situation becomes
worse. We do not display the case $\kappa_{10}=2$ for which
$\left| m_{H_{U}}\right| $ exceeds 2 TeV. On the
other hand one can see how fine tuning condition for little
hierarchy problem is improved when $\kappa=0$ at GUT scale.

In Figure \ref{1010bar2} we plot $\left| m_{H_{U}}\right| $
versus the $m_{h}$ for MSSM$+10+
\overline{10}$, with $A_{t}<0$. Solid, dashed and
dashed-dotted curve correspond to $M_S=450$, 500 and 600 GeV
respectively. For all cases $\kappa_{10}=0$ at $M_G$. We see
that the fine tuning condition is relaxed compared to the results in
Figure \ref{MSSM}.

\subsection{ MSSM$+5+\overline{5}$}

In this section we  consider  MSSM$+5+\overline{5}$ 
and  perform an analysis similar to what we did for
MSSM$+10+\overline{10}$.

\vspace{0.3cm}
{\it Case I}.  Using the boundary conditions
\begin{equation}
\{\alpha
_{G},M_{G},y_{t}(M_{G}),\kappa_{5}(M_{G})\}=\{{1}/{19.06},2.0\times
10^{16},0.57,2\},
\end{equation}
we obtain the following spectrum
\begin{eqnarray}
\{M_{1}\left( M_{Z}\right) ,M_{2}\left( M_{Z}\right) ,M_{3}\left(
M_{Z}\right) \}&=&\{0.323,0.645,2.23\}m_{1/2}
\end{eqnarray}
and
\begin{eqnarray}
M_{Z}^{2} &=&-1.26\mu
_{0}^{2}+5.32m_{1/2}^{2}+0.94 m_{0}^{2}+0.18 A_{t_{0}}^{2}+0.04 A_{\kappa _{5_{0}}}^2 \nonumber\\
&&-0.86 m_{1/2}A_{t_{0}}+0.21 m_{1/2}A_{\kappa_{5_{0}}}-0.09A_{t_{0}}A_{\kappa_{5_{0}}} 
\end{eqnarray}
\begin{eqnarray}
-m_{H_{u}}^{2}\left( M_{Z}\right)
&=&2.66m_{1/2}^{2}+0.47m_{0}^{2}+0.09A_{t_{0}}^{2}+0.02A_{\kappa _{5_0}}^{2}\nonumber \\
&&-0.43m_{1/2}A_{t_{0}}+0.1 m_{1/2}A_{\kappa _{5_{0}}}-0.044A_{t_{0}}A_{\kappa_{5_{0}}} \\
m_{Q_{t}}^{2}\left( M_{Z}\right)
&=&4.68 m_{1/2}^{2}+0.7 m_{0}^{2}-0.04 A_{t_{0}}^{2}+0.003A_{\kappa _{5_0}}^{2}\nonumber \\
&&+0.16 m_{1/2}A_{t_{0}}-0.04 m_{1/2}A_{\kappa _{5_{0}}}+0.013A_{t_{0}}A_{\kappa_{5_{0}}}\\
m_{U_{t}}^{2}\left( M_{Z}\right) &=&3.24m_{1/2}^{2}+0.39
m_{0}^{2}-0.072 A_{t_{0}}^{2}-0.006 A_{\kappa _{5_0}}^{2}
\nonumber \\
&&+0.32 m_{1/2}A_{t_{0}}-0.08 m_{1/2}A_{\kappa _{5_{0}}}+0.026
A_{t_{0}}A_{\kappa_{5_{0}}}\\
A_{t}\left( M_{Z}\right) &=&-2.17 m_{1/2}+0.28 A_{t_{0}}-0.076 A_{\kappa _{5_0}} \\
A_{\kappa _{5}}\left( M_{Z}\right) &=&0.36 m_{1/2}-0.21
A_{t_{0}}+0.16 A_{\kappa_{5_0}} \\
m_{Q_{1,2}}^{2}\left( M_{Z}\right) &=&5.73 m_{1/2}^{2}+m_{0}^{2} \\
m_{U_{1,2}}^{2}\left( M_{Z}\right) &=&5.36m_{1/2}^{2}+m_{0}^{2} \\
\mu ^{2}\left( M_{Z}\right) &=&0.63 \mu _{0}^{2} \\
m_{D_{1,2,3}}^{2}\left( M_{Z}\right) &=&m_{0}^{2}+5.31 m_{1/2}^{2} \\
m_{L_{1,2,3}}^{2}\left( M_{Z}\right) &=&m_{0}^{2}+0.474 m_{1/2}^{2} \\
m_{E_{1,2,3}}^{2}\left( M_{Z}\right) &=&m_{0}^{2}+0.141 m_{1/2}^{2}.
\end{eqnarray}
\begin{figure}[t]
\centering \includegraphics[angle=0, width=13cm]{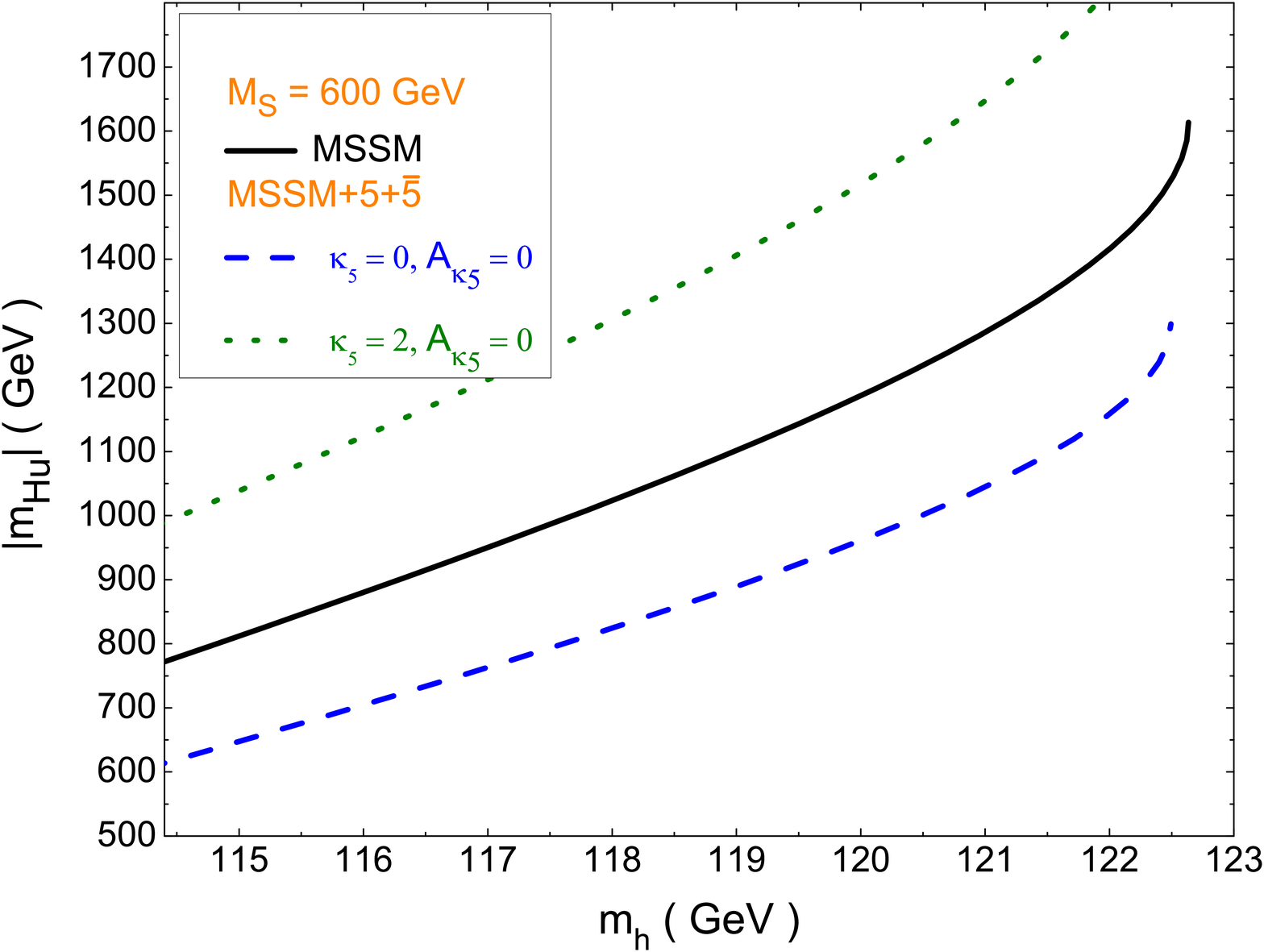}
\vspace{-1cm}
\caption{$\left| m_{H_{U}}\right| $ \ versus $m_{h}$ for MSSM$+5+%
\overline{5}$ ($A_{t}<0$).} \label{55bar1}
\end{figure}
For those particles outside the the MSSM, we find
\begin{eqnarray}
m_{L_{5}}^{2}\left( M_{Z}\right) &=&0.51 m_{1/2}^{2}+0.45
m_{0}^{2}+0.02 A_{t_{0}}^{2}-0.03A_{\kappa _{5_0}}^{2}
\nonumber \\
&&-0.042 m_{1/2}A_{t_{0}}+0.015 m_{1/2}A_{\kappa_{5_{0}}}+0.005A_{t_{0}}A_{\kappa _{5_{0}}}.
\end{eqnarray}
\begin{eqnarray}
m_{S}^{2}\left( M_{Z}\right) &=&0.073 m_{1/2}^{2}-0.11
m_{0}^{2}+0.037 A_{t_{0}}^{2}-0.06 A_{\kappa _{5_0}}^{2}
\nonumber \\
&&-0.088 m_{1/2}A_{t_{0}}+0.03 m_{1/2}A_{\kappa_{5_{0}}}+0.01 A_{t_{0}}A_{\kappa _{5_{0}}}\\
m_{D_{5}}^{2}\left( M_{Z}\right) &=&m_{0}^{2}+5.31 m_{1/2}^{2}.
\end{eqnarray}
\begin{figure}[t]
\centering \includegraphics[angle=0, width=13cm]{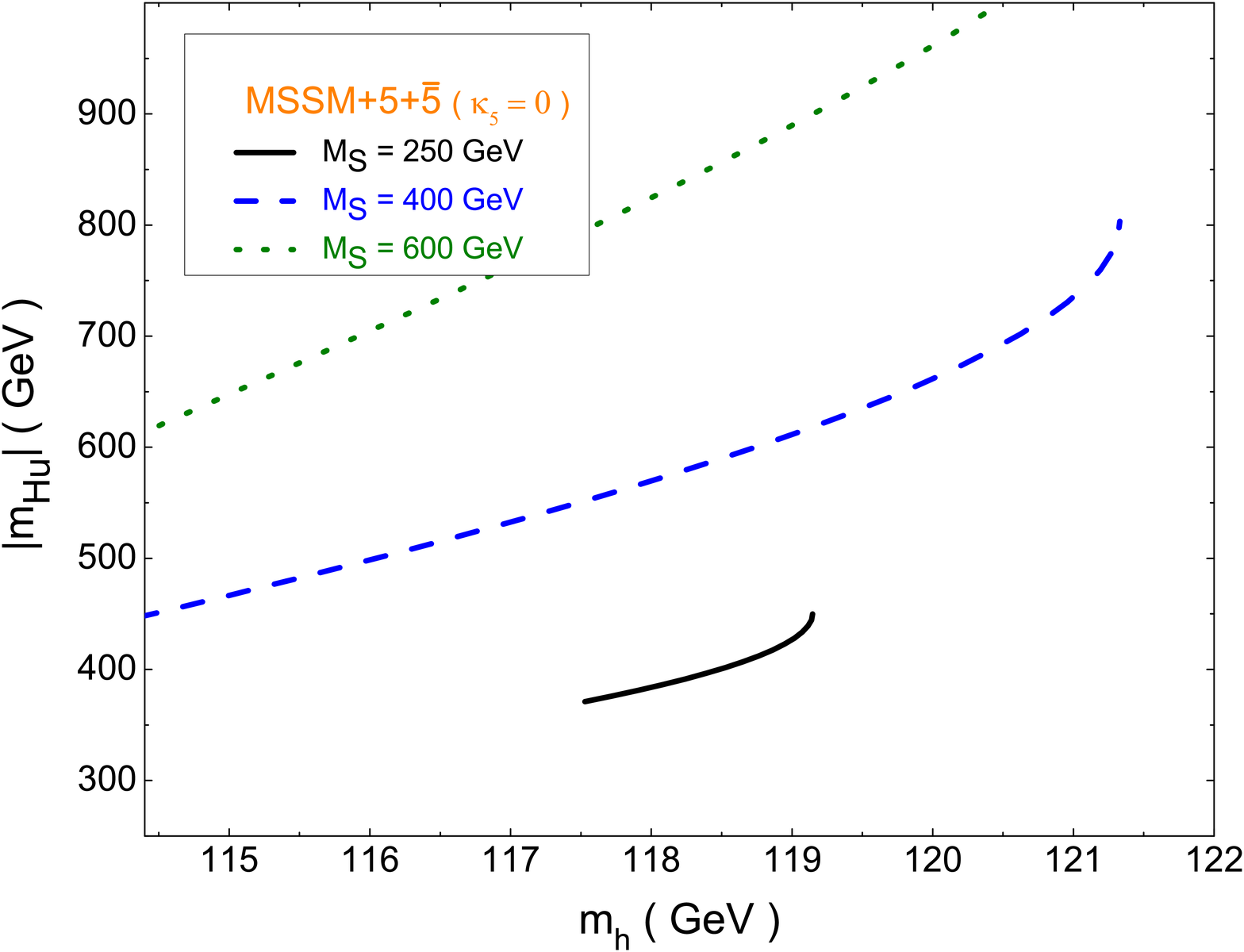}
\vspace{-1cm}
\caption{$\left| m_{H_{U}}\right| $ \ versus $m_{h}$ for MSSM$+5+%
\overline{5}$ ($A_{t}<0$)} \label{55bar2}
\end{figure}
{\it Case II}.  Employing the boundary conditions
\begin{equation}
\{\alpha
_{G},M_{G},y_{t}(M_{G}),\kappa_{5}(M_{G})\}=\{{1}/{19.06},2.0\times
10^{16},0.39,0\}.
\end{equation}
we find
\begin{equation}
\{M_{1}\left( M_{Z}\right) ,M_{2}\left( M_{Z}\right) ,M_{3}\left(
M_{Z}\right) \}=\{0.323,0.645,2.23\}m_{1/2}
\end{equation}%
and
\begin{eqnarray}
M_{Z}^{2} &=&-2.41\mu _{0}^{2}+5.37 m_{1/2}^{2}+0.03m_{0}^{2}+0.22 A_{t_{0}}^{2}-1.1 m_{1/2}A_{t_{0}}\\
m_{H_{u}}^{2}\left( M_{Z}\right)
&=&2.68 m_{1/2}^{2}+ 0.014 m_{0}^{2}+0.11 A_{t_{0}}^{2}-0.53 m_{1/2}A_{t_{0}} \\
m_{Q_{t}}^{2}\left( M_{Z}\right)
&=&4.68 m_{1/2}^{2}+0.66 m_{0}^{2}-0.04 A_{t_{0}}^{2}+0.18 m_{1/2}A_{t_{0}} \\
m_{U_{t}}^{2}\left( M_{Z}\right)
&=&3.25 m_{1/2}^{2}+0.32 m_{0}^{2}-0.073 A_{t_{0}}^{2}+0.35 m_{1/2}A_{t_{0}} \\
A_{t}\left( M_{Z}\right) &=&-2.25 m_{1/2}+0.32 A_{t_{0}} \\
A_{\kappa _{5}}\left( M_{Z}\right) &=&0.23 m_{1/2}-0.34
A_{t_{0}}+A_{\kappa_{5_{0}}}\\
m_{Q_{1,2}}^{2}\left( M_{Z}\right) &=&5.73 m_{1/2}^{2}+m_{0}^{2} \\
m_{U_{1,2}}^{2}\left( M_{Z}\right) &=&5.36m_{1/2}^{2}+m_{0}^{2} \\
\mu ^{2}\left( M_{Z}\right) &=&1.20 \mu _{0}^{2} 
\end{eqnarray}
\begin{eqnarray}
m_{D_{1,2,3}}^{2}\left( M_{Z}\right) &=&m_{0}^{2}+5.31 m_{1/2}^{2} \\
m_{L_{1,2,3}}^{2}\left( M_{Z}\right) &=&m_{0}^{2}+0.474 m_{1/2}^{2} \\
m_{E_{1,2,3}}^{2}\left( M_{Z}\right) &=&m_{0}^{2}+0.141 m_{1/2}^{2}
\end{eqnarray}%
Similarly,
\begin{eqnarray}
m_{L_{5}}^{2}\left( M_{Z}\right)
&=&0.47 m_{1/2}^{2}+ m_{0}^{2} \\
m_{S}^{2}\left( M_{Z}\right)&=& m_{0}^{2} \\
m_{D_{5}}^{2}\left( M_{Z}\right) &=&m_{0}^{2}+5.31 m_{1/2}^{2}.
\end{eqnarray}

In Figure \ref{55bar1} we plot  $\left| m_{H_{U}}\right| $ versus
$m_{h}$, with $M_S=600$ GeV. We consider
two cases, $\kappa_5=2$ and $ \kappa_5=0$ at $M_G$. We can see
from Figure \ref{55bar1} that the fine tuning condition for the little
hierarchy problem improves relative  to  MSSM when $ \kappa_5=0$,
but improvement is not as significant as for MSSM$+10+\overline{10}$.
The reason is that the values for $y_t$ at $M_G$
with $\kappa_5 = 0$ is large compared to $y_t$ for
MSSM$+10+\overline{10}$ with $\kappa_{10}=0$.

In Figure \ref{55bar2} we plot $\left| m_{H_{U}}\right| $ versus
$m_{h}$ for MSSM$+5+ \overline{5}$, with ($A_{t}<0$). Solid, dashed 
and dashed-dotted curve correspond to $M_S=250$, 400 and 600 GeV 
respectively, with $\kappa_{5}=0$ at $M_G$. We see that the fine 
tuning condition is relaxed compared to the results in Figure \ref{MSSM}.

\section{Conclusion}

 We have shown that in an extended MSSM framework with
vectorlike supermultiplets, whose masses lie in the TeV range, the
mass of the lightest CP even Higgs boson can be as high as $160$
GeV. Gauge coupling unification is maintained in this approach with
the three MSSM gauge couplings remaining perturbative all the way to
the GUT scale $M_G$. As far as the little hierarchy problem is
concerned, the degree of fine tuning in this extended MSSM is 
severe for larger values of the Higgs mass. However, things have 
improved somewhat compared to the MSSM if the Higgs mass is 
found to be $\lesssim 125$ GeV.

\section{Acknowledgements}
We thank Christopher Kolda for collaboration in the early stages of
this work and for discussion. This work is supported in part by the DOE
under grant \# DE-FG02-91ER40626 (I.G. and Q.S.), \# DE-FG03-98ER-41076 (K.S.B.), 
and GNSF grant 07\_462\_4-270 (I.G.). 

%
%
%

\section{Appendix}
\begin{appendix}

\section{RGEs for MSSM$+10+\overline{10}$}

\begin{eqnarray*}
W &=&y_{t}Q_{t}\overline{U_{t}}H_{u}+\kappa _{10}Q_{10}\overline{U}_{10}H_{u}
\\
\frac{d\overline{\alpha }_{i}}{dt} &=&-b_{i}\overline{\alpha }_{i}^{2} \\
\frac{dM_{i}}{dt} &=&-b_{i}\overline{\alpha }_{i}M_{i} \\
\text{Here }t &=&\log \left( \frac{M_{G}^{2}}{Q^{2}}\right) ,\overline{%
\alpha }_{i}=\frac{\alpha _{i}}{4\pi } \\
\frac{dm_{L}^{2}}{dt} &=&\left( 3\overline{\alpha }_{2}M_{2}^{2}++\frac{3}{5}%
\overline{\alpha }_{1}M_{1}^{2}\right)  \\
\frac{dm_{E}^{2}}{dt} &=&\left( \frac{12}{5}\overline{\alpha }%
_{1}M_{1}^{2}\right)  \\
\frac{dm_{Q_{t}}^{2}}{dt} &=&\left( \frac{16}{3}\overline{\alpha }%
_{3}M_{3}^{2}+3\overline{\alpha }_{2}M_{2}^{2}+\frac{1}{15}\overline{\alpha }%
_{1}M_{1}^{2}\right)  \\
&&-Y_{t}\left( m_{Q_{t}}^{2}+m_{U_{t}}^{2}+m_{H_{u}}^{2}+A_{t}^{2}\right)  \\
\frac{dm_{U_{t}}^{2}}{dt} &=&\left( \frac{16}{3}\overline{\alpha }%
_{3}M_{3}^{2}+\frac{16}{15}\overline{\alpha }_{1}M_{1}^{2}\right)
-2Y_{t}\left( m_{Q_{t}}^{2}+m_{U_{t}}^{2}+m_{H_{u}}^{2}+A_{t}^{2}\right)  \\
\frac{d\mu ^{2}}{dt} &=&3\left( \overline{\alpha }_{2}+\frac{1}{5}\overline{%
\alpha }_{1}\right) \mu ^{2}-3\left( Y_{t}+K_{10}\right) \mu ^{2} \\
\frac{dm_{H_{d}}^{2}}{dt} &=&3\left( \overline{\alpha }_{2}M_{2}^{2}+\frac{1%
}{5}\overline{\alpha }_{1}M_{1}^{2}\right)  \\
\frac{dm_{H_{u}}^{2}}{dt} &=&3\left( \overline{\alpha }_{2}M_{2}^{2}+\frac{1%
}{5}\overline{\alpha }_{1}M_{1}^{2}\right) -3Y_{t}\left(
m_{Q_{t}}^{2}+m_{U_{t}}^{2}+m_{H_{u}}^{2}+A_{t}^{2}\right)  \\
&&-3K_{10}\left( m_{Q_{10}}^{2}+m_{U_{10}}^{2}+m_{H_{u}}^{2}+A_{\kappa
_{10}}^{2}\right)  \\
\frac{dA_{t}}{dt} &=&-\left( \frac{16}{3}\overline{\alpha }_{3}M_{3}+3%
\overline{\alpha }_{2}M_{2}+\frac{13}{15}\overline{\alpha }_{1}M_{1}\right)
-6Y_{t}A_{t}-3K_{10}A_{\kappa_{10} } \\
\frac{dY_{t}}{dt} &=&Y_{t}\left( \frac{16}{3}\overline{\alpha }_{3}+3%
\overline{\alpha }_{2}+\frac{13}{15}\overline{\alpha }_{1}\right)
-6Y_{t}^{2}-3Y_{t}K_{10} \\
\frac{dm_{Q_{10}}^{2}}{dt} &=&\left( \frac{16}{3}\overline{\alpha }%
_{3}M_{3}^{2}+3\overline{\alpha }_{2}M_{2}^{2}+\frac{1}{15}\overline{\alpha }%
_{1}M_{1}^{2}\right)  \\
&&-K_{10}\left( m_{Q_{10}}^{2}+m_{U_{10}}^{2}+m_{H_{u}}^{2}+A_{\kappa
_{10}}^{2}\right)  \\
\frac{dm_{U_{10}}^{2}}{dt} &=&\left( \frac{16}{3}\overline{\alpha }%
_{3}M_{3}^{2}+\frac{16}{15}\overline{\alpha }_{1}M_{1}^{2}\right) -2K_{10}\left(
m_{Q_{10}}^{2}+m_{U_{10}}^{2}+m_{H_{u}}^{2}+A_{\kappa_{10} }^{2}\right)  \\
\frac{dA_{\kappa_{10} }}{dt} &=&-\left( \frac{16}{3}\overline{\alpha }_{3}M_{3}+3%
\overline{\alpha }_{2}M_{2}+\frac{13}{15}\overline{\alpha }_{1}M_{1}\right)
-6K_{10}A_{\kappa_{10} }-3Y_{t}A_{t} 
\end{eqnarray*}
\begin{eqnarray*}
\frac{dK_{10}}{dt} &=&K_{10}\left( \frac{16}{3}\overline{\alpha }_{3}+3\overline{%
\alpha }_{2}+\frac{13}{15}\overline{\alpha }_{1}\right) -6K_{10}^{2}-3Y_{t}K_{10} \\
\frac{dm_{D}^{2}}{dt} &=&\left( \frac{16}{3}\overline{\alpha }_{3}M_{3}^{2}+%
\frac{4}{15}\overline{\alpha }_{1}M_{1}^{2}\right). \\
\text{Here }Y_{t} &=&\frac{y_{t}^{2}}{\left( 4\pi \right) ^{2}},\text{ }K_{10}=%
\frac{\kappa _{10}^{2}}{\left( 4\pi \right) ^{2}},b_{i}=\left\lbrace \frac{33}{5},1,-3\right\rbrace +\left\lbrace 3,3,3\right\rbrace .
\end{eqnarray*}

\section{RGEs for MSSM$+5+\overline{5}$}

\begin{eqnarray*}
W &=&y_{t}Q_{t}\overline{U_{t}}H_{u}+\kappa _{5}L_{5}\overline{S}H_{u} \\
\frac{dm_{L}^{2}}{dt} &=&3\left( \overline{\alpha }_{2}M_{2}^{2}+\frac{1}{5}%
\overline{\alpha }_{1}M_{1}^{2}\right)  \\
\frac{dm_{E}^{2}}{dt} &=&\left( \frac{12}{5}\overline{\alpha }%
_{1}M_{1}^{2}\right)  \\
\frac{dm_{Q_{t}}^{2}}{dt} &=&\left( \frac{16}{3}\overline{\alpha }%
_{3}M_{3}^{2}+3\overline{\alpha }_{2}M_{2}^{2}+\frac{1}{15}\overline{\alpha }%
_{1}M_{1}^{2}\right)  \\
&&-Y_{t}\left( m_{Q_{t}}^{2}+m_{U_{t}}^{2}+m_{H_{u}}^{2}+A_{t}^{2}\right)  \\
\frac{dm_{U_{t}}^{2}}{dt} &=&\left( \frac{16}{3}\overline{\alpha }%
_{3}M_{3}^{2}+\frac{16}{15}\overline{\alpha }_{1}M_{1}^{2}\right)
-2Y_{t}\left( m_{Q_{t}}^{2}+m_{U_{t}}^{2}+m_{H_{u}}^{2}+A_{t}^{2}\right)  \\
\frac{d\mu ^{2}}{dt} &=&3\left( \overline{\alpha }_{2}+\frac{1}{5}\overline{%
\alpha }_{1}\right) \mu ^{2}-\left( 3Y_{t}+K_{5}\right) \mu ^{2} \\
\frac{dm_{H_{d}}^{2}}{dt} &=&3\left( \overline{\alpha }_{2}M_{2}^{2}+\frac{1%
}{5}\overline{\alpha }_{1}M_{1}^{2}\right)  \\
\frac{dm_{H_{u}}^{2}}{dt} &=&3\left( \overline{\alpha }_{2}M_{2}^{2}+\frac{1%
}{5}\overline{\alpha }_{1}M_{1}^{2}\right) -3Y_{t}\left(
m_{Q_{t}}^{2}+m_{U_{t}}^{2}+m_{H_{u}}^{2}+A_{t}^{2}\right)  \\
&&-K_{5}\left( m_{L_{5}}^{2}+m_{S}^{2}+m_{H_{u}}^{2}+A_{\kappa
_{5}}^{2}\right)  \\
\frac{dA_{t}}{dt} &=&-\left( \frac{16}{3}\overline{\alpha }_{3}M_{3}+3%
\overline{\alpha }_{2}M_{2}+\frac{13}{15}\overline{\alpha }_{1}M_{1}\right)
-6Y_{t}A_{t}-K_{5}A_{\kappa _{5}} \\
\frac{dY_{t}}{dt} &=&Y_{t}\left( \frac{16}{3}\overline{\alpha }_{3}+3%
\overline{\alpha }_{2}+\frac{13}{15}\overline{\alpha }_{1}\right)
-6Y_{t}^{2}-Y_{t}K_{5} \\
\frac{dm_{L_{5}}^{2}}{dt} &=&3\left( \overline{\alpha }_{2}M_{2}^{2}+\frac{1%
}{5}\overline{\alpha }_{1}M_{1}^{2}\right) -K_{5}\left(
m_{L_{5}}^{2}+m_{S}^{2}+m_{H_{u}}^{2}+A_{\kappa _{5}}^{2}\right)  \\
\frac{dm_{S}^{2}}{dt} &=&-2K_{5}\left(
m_{L_{5}}^{2}+m_{S}^{2}+m_{H_{u}}^{2}+A_{\kappa _{5}}^{2}\right)  \\
\frac{dA_{\kappa _{5}}}{dt} &=&-3\left( \overline{\alpha }_{2}M_{2}+\frac{1}{%
5}\overline{\alpha }_{1}M_{1}\right) -4K_{5}A_{\kappa _{5}}-3Y_{t}A_{t} \\
\frac{dK_{5}}{dt} &=&3K_{5}\left( \overline{\alpha }_{2}+\frac{1}{5}%
\overline{\alpha }_{1}\right) -4K_{5}^{2}-3Y_{t}K_{5} 
\end{eqnarray*}
\begin{eqnarray*}
\frac{dm_{D}^{2}}{dt} &=&\left( \frac{16}{3}\overline{\alpha }_{3}M_{3}^{2}++%
\frac{4}{15}\overline{\alpha }_{1}M_{1}^{2}\right).  \\
\text{Here }Y_{t} &=&\frac{y_{t}^{2}}{\left( 4\pi \right) ^{2}},\text{ }%
K_{5}=\frac{\kappa _{5}^{2}}{\left( 4\pi \right) ^{2}},b_{i}=\left\lbrace \frac{33}{5},1,-3\right\rbrace +\left\lbrace 1,1,1\right\rbrace .
\end{eqnarray*}
\end{appendix}


\end{document}